\pdfoutput=1

\documentclass[twocolumn,floatfix,twocolappendix,times,reprint]{aastex631}

\usepackage{graphicx} 
\graphicspath{{Figures/}} 

\usepackage{dcolumn} 
\usepackage{bm} 
\usepackage{float} 
\usepackage{booktabs} 
\usepackage{enumitem} 

\usepackage{amsmath} 
\usepackage{amssymb} 
\usepackage{latexsym}

\usepackage{cleveref}

\usepackage[english]{babel}
\usepackage[expansion=all,babel]{microtype}
\usepackage{mathtools}

\usepackage{multirow}
\usepackage[flushmargin]{footmisc} 

\newcommand{\Msun}{\mathrm{M}_\odot}

\newcommand{\rhob}{\rho_\mathrm{b}}
\newcommand{\rhoc}{\rho_\mathrm{c}}
\newcommand{\Ts}{T_\mathrm{s}}
\newcommand{\Tb}{T_\mathrm{b}}

\newcommand{\dd}{\mathrm{d}}

\newcommand{\gcc}{\mathrm{g~cm^{-3}}}
\newcommand{\slfrac}[2]{\left.#1\middle/#2\right.}
\newcommand{\e}{\mathrm{e}}
\newcommand{\half}{\frac{1}{2}}

\makeatletter
	\newcommand{\vast}{\bBigg@{3.4}}
\makeatother

\submitjournal{ApJ}

\shorttitle{Standard cooling of rotating neutron stars in 2D} 
\shortauthors{Beznogov et al.}

\begin{document}

\title{Standard cooling of rapidly rotating isolated neutron stars in 2D}
\author[0000-0002-7326-7270]{Mikhail V. Beznogov}
\affiliation{National Institute for Physics and Nuclear Engineering (IFIN-HH), RO-077125 Bucharest, Romania}
\affiliation{Instituto de Astronom\'ia, Universidad Nacional Aut\'onoma de M\'exico, Ciudad de M\'exico, CDMX 04510, Mexico}
\email{mikhail.beznogov@nipne.ro}

\author[0000-0002-6029-4712]{J\'er\^ome Novak}
\affiliation{Laboratoire Univers et Th\'eories, Observatoire de Paris, Universit\'e PSL, CNRS, Universit\'e de Paris-Cit\'e, 92190 Meudon, France}
\email{jerome.novak@obspm.fr}

\author[0000-0003-2498-4326]{Dany Page}
\affiliation{Instituto de Astronom\'ia, Universidad Nacional Aut\'onoma de M\'exico, Ciudad de M\'exico, CDMX 04510, Mexico}
\email{page@astro.unam.mx}

\author[0000-0001-8421-2040]{Adriana R. Raduta}
\affiliation{National Institute for Physics and Nuclear Engineering (IFIN-HH), RO-077125 Bucharest, Romania}
\email{araduta@nipne.ro}

\date{\today}
             
\begin{abstract}
We study the long-term thermal evolution of axisymmetric rotating neutron stars in full general relativity. To this aim, we develop \texttt{NSCool 2D Rot}, a major upgrade of the 1D neutron stars thermal evolution code \texttt{NSCool} by \cite{NSCool}. As a first application of our new code we address the standard cooling of isolated neutron stars with rotation frequencies up to the mass shedding limit. We investigate the effects of the equation of state (EOS) by considering different combinations of core and crust EOSs. The results indicate complex time-dependent evolution of temperature distribution throughout the whole volume of the star and, in particular, in the crust. We show that most of that complexity can be attributed to the formation of a ``heat blob'' in the crust and to the latitude dependence of the heat diffusion timescale through the crust. 
\end{abstract}

\keywords{dense matter--stars: neutron} 
\section{Introduction}
\label{sec:Intro}

Neutron stars (NSs) are born in core-collapse supernova events \citep{Baade_1934} or in accretion-induced collapses of white dwarfs \citep{Canal_1976} and start their lives as proto-neutron stars \citep{Burrows_1986}. Alternatively, a NS can be ``re-born'' in the merger of a binary NSs system provided that it survives the event (i.e., does not collapse into a black hole, see \citealt{Kluzniak_1998}). In a core-collapse supernova, the proto-neutron star phase starts after the core bounce and ends a few tens of seconds later when a proto-neutron star becomes an ``ordinary'' NS that is transparent to neutrinos. By this time the deleptonization process is already finished, the entropy has decreased significantly (see, e.g., \citealt{Pons_1999,Roberts_2012,Nagakura_2020,Pascal_2022}) and the star has (almost) reached its final size and composition. During the next few hours to a day it will cool down to temperatures well below $10^{10}$~K, shrink a little bit more and form a crust \citep{Beznogov_2020}. After the end of this so-called neo-neutron star phase a NS can be considered as composed of cold catalyzed matter.

NSs provide a unique opportunity to study the properties of cold superdense matter (e.g., \citealt{HPY07}), its equation of state (EOS) being the ``most wanted'' characteristic. The mystery of the EOS involves two facets: pressure ($P$) vs density ($\rho$) dependence and chemical composition. The former can be inferred by different techniques such as, e.g., independent measurements of NS masses and radii (e.g., \citealt{Miller_2019,Miller_2021,Riley_2019,Riley_2021}) and gravitational wave (GW) observations of binary NS mergers (e.g., \citealt{Abbott_2017,Abbott_2019}). The question of composition, however, is more elusive and the study of NS thermal evolution is one of the most promising tools to tackle it.

To investigate thermal evolution one needs to solve two types of (in the general case) coupled equations: mechanical structure equations and thermal transport \& balance equations. However, in the particular case of the \emph{long-term} evolution the situation gets significantly simplified. As mentioned earlier, within a few hours to a day after the birth of a NS its temperature drops well below $10^{10}$~K. At such temperatures thermal effects on NS composition and mechanical structure become negligible everywhere, except for the outermost layers of the star which constitute the heat blanketing envelope and have a negligible impact on the mechanical properties of the rest of the star \citep{Baym_1981}. Thus, the chemical composition and equilibrium configurations may be assessed by employing a barotropic EOS of cold baryonic matter in $\beta$-equilibrium everywhere, except for the outer envelope. This allows one to decouple mechanical structure and thermal evolution equations. The former then has to be calculated only once while the latter are solved repeatedly as time progresses (e.g., \citealt{Baym_1981,HPY07,PPP15}). In this situation the surface temperature, which is the only really ``measurable'' quantity, is determined by the heat blanketing envelope via the so-called ``$\Ts-\Tb$'' relation (e.g., \citealt{GPE83,Beznogov_2021}). The validity of such an approach to the long-term cooling of NSs was explicitly demonstrated by \cite{Dong_2018} for the spherically symmetric (1D) case by comparing the cooling of NSs  with dynamic (i.e., non-fixed) and static (i.e., fixed) mechanical structures. For more subtle details of this widely used method see Appendix~\ref{App:Envelope}. Throughout this work only the ``long-term thermal evolution (or cooling)'' will be addressed. Though for convenience sometimes we shall call it simply ``thermal evolution (or cooling)".

Under the hypothesis that no source of internal heating is present, the thermal history depends upon neutrino emission mechanisms, transport properties and superfluidity gaps of various particle degrees of freedom \citep{YKGH01,YP04,PGW06,Page:2009kx,Tsuruta_2009}, all of which being strongly dependent on the EOS \citep{Raduta_MNRAS_2018,Fortin_PRD_2021}. Stars accommodating at least one direct Urca (DURCA) process cool much faster than stars whose thermal evolution is driven by the series of the so-called slow (modified Urca and bremsstrahlung) and, in the case of paired species, intermediate (Cooper pair breaking and formation) cooling processes. Whether DURCA processes occur or not depends upon the density dependence of the nuclear symmetry energy; nucleation of heavy baryons and quarks; condensations of pions and kaons.  Whether one or more species are paired depends upon the magnitude and density dependence of the pairing gaps of fermions, both of them controlled by the baryon interactions and the many particle correlations accounted for. Uncertainties in nuclear physics beyond the saturation density of isospin symmetric matter render the appearance of fast and intermediate cooling processes and, consequently, the thermal evolution highly model dependent. 

Thermal evolution equations suitable for studying axisymmetric NSs with arbitrary rotation frequencies were derived from first principles by \cite{Miralles_1993}, who also investigated the cooling of rotating isolated NSs. In that work, the mechanical structure of the star was computed within the slow rotation approximation \citep{Hartle_1967,Hartle_1968}, while the isothermal interior approximation was used for the thermal evolution. The latter assumption means that the redshifted temperature is uniform throughout the volume, which makes it not directly affected by rotation. In such a situation the angular dependence of surface temperature exists only due to the dependence of heat blanketing envelope solution on surface gravitational acceleration \citep{GPE82}. It was also shown, albeit on a schematic model, that the cold wave from the core reaches the equator later than the poles (the core is usually cooling down faster than the crust due to more intensive neutrino emission; hence the cold wave).

More sophisticated investigations were done by \cite{Schaab_1998}, who improved upon both mechanical structure and thermal evolution calculations. Their results confirmed the cooling wave propagation conclusions of \cite{Miralles_1993} and also demonstrated that fast rotation delays the crust relaxation phase by a few hundred years.

Cooling of NSs rotating at a constant frequency was later on considered by \cite{Negreiros_2012}. When the constant frequency assumption is relaxed, e.g., by taking into account the pulsar spin-down, modifications of microscopic properties affect the cooling history \citep{Negreiros_2013}. Considering that the DURCA process is only allowed above a critical density, $\rho_\mathrm{DU}$, whose value is EOS dependent, a fast rotating NS whose central density $\rho_c$ was initially below $\rho_\mathrm{DU}$ may see, due to the decrease of the centrifugal force during the spin-down, $\rho_c$ increase sufficiently to eventually surpass $\rho_\mathrm{DU}$ and slowly turn-on the DURCA process in a time growing inner core. Simulations accounting for fully time-dependent microphysics, including superfluid gaps, and mechanical structure were performed within the ``thermo-rotational'' model by \citet{Negreiros_2017}.

A common feature of these works is that they focus on the surface temperature distribution, which is strongly affected by the value of local surface gravitational acceleration. The figures in \citet{Negreiros_2012,Negreiros_2017} nevertheless suggest that internal temperature distributions are worth more attention.

Thermal evolution of NSs in 2D was also investigated in relation with magnetic fields and/or extra heat sources in the crust. Magneto-thermal evolution works usually assuming a spherically symmetric (1D) mechanical structure and an axisymmetric (2D) thermal and magnetic structure (e.g., \citealt{Vigano_2021}). Note that ``mixed'' 1D \& 2D calculations, but this time for the core and the crust, respectively, have been employed for magnetized NSs \citep{Geppert_2004} as well as for extra heat sources in the crust \citep{KKPY14}.

Our project aims to generalize the studies of long-term thermal evolution of NSs beyond spherical symmetry. This will allow us to more realistically investigate rotating isolated NSs; accreting NSs which are heated anisotropically; NSs deformed under strong magnetic fields. In order to account as much as possible for the rich microphysics relevant for these topics and take benefit from the existing expertise in the field we made the choice of upgrading \texttt{NSCool} \citep{NSCool}, a code in 1D, extensively used for thermal evolution simulations of a large variety of NSs among which are long-term cooling of isolated stars \citep{PLPS04,Page_etal09}, cooling of neo-neutron stars \citep{Beznogov_2020}, heating and cooling of accreting stars in low mass X-ray binaries \citep{Ootes_2016,Degenaar_2021,Page_2022}. In the present paper we shall consider only nucleonic rigidly rotating NSs in the standard long-term cooling paradigm.
 
The paper is organized as follows. In Sect.~\ref{sec:ThEqs} we briefly present the spacetime metric appropriate for describing rigidly rotating NSs as well as the equations of their thermal evolution. Models of EOSs and the mechanical structure of NSs rotating up to the Kepler limit are considered in Sect.~\ref{sec:Models}. In Sect.~\ref{sec:Cooling} we present the results of our cooling simulations. The conclusions are presented in Sect.~\ref{sec:Concl}. Throughout this paper by matter density ($\rho$) we mean energy density including rest-mass divided by the speed of light squared.  

\section{Equations for thermal evolution of rotating stars} 
\label{sec:ThEqs}

\subsection{Spacetime metric}
\label{ssec:metric}

Calculation of the space-time metric and structure of axisymmetric rigidly rotating NS in full general relativity (GR) is a complicated issue. Relatively simple procedures exist only in the limit of slow rotation, where rotation induced modifications can be treated as small perturbations of an (already known) static configuration~\citep{Hartle_1967,Hartle_1968}. For rapidly rotating NSs more sophisticated techniques are required for solving the coupled system of mechanical structure and Einstein field equations. One of the most efficient and accurate methods was proposed by \cite{Bonazzola_1993}; the numerical integration is based on a spectral method. In the present work we employ the \texttt{LORENE} library \footnote{\url{https://lorene.obspm.fr}} \citep{LORENE_2016}, suitable to deal with rotation frequencies from zero up to the mass shedding limit.

Herein we briefly review the working hypothesis; choice of coordinates; metric potentials and NS mechanical structure; for more information, see \cite{Bonazzola_1993}.

The motion of matter is assumed to be 
(i) axisymmetric [with respect to the rotational axis],
(ii) stationary, 
(iii) purely azimuthal [i.e. no convection, no meridional currents; we also neglect all mass flows induced by heat transport]. 
And
(iv) the EOS of the matter is barotropic. For possible relaxation of (iii) see, e.g., \cite{Birkl_2011} and of (iv) see, e.g., \cite{Camelio_2019}.

Following \cite{Bonazzola_1993}, we use the \emph{quasi-isotropic} gauge, relying on spherical coordinates with the polar angle $\theta$ measured from the rotation axis. Note that the quasi-isotropic coordinates used in \texttt{LORENE} are different from the conventional Schwarzschild coordinates used in the spherically symmetric Tolman–Oppenheimer–Volkoff equations\,\footnote{In spherical symmetry, the quasi-isotropic gauge turns into the isotropic gauge, for which there exists a coordinate transform to the Schwarzschild one. Note, however, that outside spherical symmetry there is no gauge generalizing the Schwarzschild coordinates.}.   

To write down the components of the metric tensor we need to additionally specify the choice of the basis tetrad (basis vectors). Choosing ``natural'' tetrad $\bm{e}_\alpha = \partial_\alpha = \slfrac{\partial}{\partial x^\alpha}$ with $x^\alpha = \left(c\,t,r,\theta,\varphi \right)^\mathrm{T}$, we obtain the following expression in the notation of \cite{Gourgoulhon_1999}\,\footnote{In \citep{Bonazzola_1993} some components of the metric tensor are defined differently, while the notation of \cite{Gourgoulhon_1999} corresponds exactly to the one used in \texttt{LORENE}.}:
\begin{widetext}
	\begin{equation}
		g_{\mathit{\alpha \beta}} = 
		\begin{pmatrix}
			-N^2 + B^2 \left(\frac{N^\varphi}{c} \right)^2 r^2 \sin^2 \theta & 0 & 0 & -B^2 \left( \frac{N^\varphi}{c} \right) r^2 \sin^2 \theta \\	
			0 & A^2 & 0 & 0 \\
			0 & 0 & A^2 r^2 & 0 \\
			-B^2 \left( \frac{N^\varphi}{c} \right) r^2 \sin^2 \theta & 0 & 0 & B^2 r^2 \sin^2 \theta
		\end{pmatrix},
		\label{Eq:full_metric}
	\end{equation}
\end{widetext}
where $N$, $A$, $B$ and $N^\varphi$ are four metric functions that depend on $r$ and $\theta$ but, because of stationarity and axisymmetry, do not depend on $t$ and $\varphi$; $c$ is the speed of light; $\alpha,\beta = 0,1,2,3$. Note that $N$, $A$, $B$ are dimensionless while $N^\varphi$ has the dimension of frequency. The line element reads
\begin{align}
		ds^2 =  g_{\mathit{\alpha \beta}} \dd x^\alpha \dd x^\beta = -&N^2 c^2 \dd t^2 + A^2 \left(\dd r^2 + r^2 \dd \theta^2 \right) + \nonumber \\
		&B^2 r^2 \sin^2 \theta \left(\dd \varphi - N^\varphi \dd t \right)^2  .
	\label{Eq:interval}
\end{align}
%

\subsection{Heat transport equations}
\label{ssec:transport}

Energy balance and energy transport equations for rotating axisymmetric systems in GR were derived by \cite{Miralles_1993} by means of the relativistic Boltzmann equation. Here we write them down following \cite{Schaab_1998}, with the notation adjusted to match \cite{Gourgoulhon_1999} [see the previous subsection; cf. also \cite{Negreiros_2012}].

The energy balance (or conservation) equation is\,\footnote{For neutrino losses $Q_\nu < 0$, the sign before the source term is chosen accordingly.}:
\begin{equation}
	\begin{split}
		\partial_r \widetilde{h}_{(r)} + \frac{1}{r}\partial_\theta \widetilde{h}_{(\theta)} =
		&-r^2\sin \theta A^2 B \Gamma \times \\
		&\left[-\left(\frac{N}{\Gamma}\right)^2 Q + C_\mathrm{V} \partial_t \widetilde{T}\right]
	\end{split},
	\label{Eq:energy_balance_1}
\end{equation}
where $C_\mathrm{V}$ is the specific heat, $Q$ stands for the heat sources/sinks and
\begin{align}
	U &= B r \sin \theta \left(\Omega' - N^\varphi \right)/N,  
	\\
	\Gamma &= \frac{1}{\sqrt{1-\left(\frac{U}{c}\right)^2}},
	\\
	\widetilde{h}_{(i)} &= \frac{A B N^2}{\Gamma} r^2 \sin \theta \, h^i, \,\, i = r, \theta, 
	\label{Eq:flux_red-shift}
	\\
	\widetilde{T} &= \frac{N}{\Gamma} T,
	\label{Eq:T_red-shift}
\end{align}
with $T$ being the local temperature and $\Omega'$ being the rotation frequency (in radians per second; $\Omega' = 2 \pi \Omega$). Redshifted quantities are denoted with a tilde. In Eq.~\eqref{Eq:flux_red-shift} the brackets in the components of the redshifted ``flux'', $\widetilde{h}_{(i)}$, emphasize that these do not form a vector. In contrast, the components of the local flux $h^i$ in the same equation do form a vector and this equation defines contravariant components, as indicated by the superscript. Also note that the local flux is defined in the comoving reference frame.

The energy flux is calculated in the diffusion approximation:
\begin{align}
	\widetilde{h}_{(r)}      &= -r^2 \sin \theta\, \mathrm{K} N B \partial_r \widetilde{T},
	\label{Eq:flux_r_def} 
	\\
	\widetilde{h}_{(\theta)} &= -r \sin \theta\, \mathrm{K} N B \partial_\theta \widetilde{T},
	\label{Eq:flux_th_def}
\end{align}
where $\mathrm{K}$ is the thermal conductivity; in the present work we assumed that $\mathrm{K}$ is isotropic (scalar).

Eqs.~\eqref{Eq:energy_balance_1}, \eqref{Eq:flux_r_def} and \eqref{Eq:flux_th_def} can be combined into a single second order partial differential equation for the evolution of the redshifted temperature $\widetilde{T}$:
\begin{align}
	\begin{split}
		\partial_t \widetilde{T} = &\frac{1}{r^2 \sin \theta A^2 B \Gamma C_\mathrm{V}} 
		\bigg[\partial_r \left(r^2 \sin \theta \mathrm{K} N B \partial_r \widetilde{T}\right) +  \\		
		&\partial_\theta \left(\sin \theta \mathrm{K} N B \partial_\theta \widetilde{T} \right) \bigg]
		+\left(\frac{N}{\Gamma}\right)^2 \frac{Q}{C_\mathrm{V}}.	
	\end{split}
	\label{Eq:Ttilde}
\end{align}
Eq.~\eqref{Eq:Ttilde} can be further recast as
\begin{align}
	\partial_t \widetilde{T} &= G_\mathit{(rr)} + G_\mathit{(\theta \theta)} + \left(\frac{N}{\Gamma}\right)^2 \frac{Q}{C_\mathrm{V}}, 
	\label{Eq:T_main} \\
	G_\mathit{(rr)} &= \frac{1}{r^2 A^2 B \Gamma C_\mathrm{V}} \partial_r \left(r^2 \mathrm{K} N B \partial_r \widetilde{T}\right), 
	\label{Eq:Grr} \\
	G_\mathit{(\theta \theta)} &= \frac{1}{r^2 \sin \theta A^2 B \Gamma C_\mathrm{V}}  \partial_\theta \left(\sin \theta \mathrm{K} N B \partial_\theta \widetilde{T} \right).
	\label{Eq:Gthth}
\end{align}

Eq.~\eqref{Eq:T_main} is our main equation. It has the form of a ``classical'' heat diffusion equation. It is parabolic and allows for instantaneous heat propagation. One may question the applicability of such an equation in the framework of GR. The issue has been previously addressed by \cite{Dommes_2020}. According to these authors, instantaneous propagation of heat and GR are not incompatible as long as the hydrodynamic description is applicable. The NSs we are interested in, i.e. with thermal evolution timescales much longer than hydrodynamic timescales, fall into this category, which makes the use of Eq.~\eqref{Eq:T_main} safe.

The solution of Eq.~\eqref{Eq:T_main} requires initial and boundary conditions, which we comment on in the following. 

\subsection{Initial and boundary conditions} 
\label{ssec:boundcond}

The initial temperature profile at the initial time is chosen, for all NS, as follows. Matter in the center of the star and at the bottom of the envelope is given redshifted temperatures of $1.5 \times 10^{10}$~K and $10^{10}$~K, respectively. In the intermediate shells $\widetilde{T}$ is assumed to scale with the logarithm of density. Such a setup corresponds roughly to constant internal flux and improves numerical convergence of the first timestep as it makes the initial and boundary conditions ``more compatible''; starting with constant redshifted temperature would have meant zero initial internal flux which is incoherent with non-zero initial surface flux\,\footnote{For neo-NS modeling a dedicated matching procedure is required to have fully compatible initial and boundary conditions, see \citep{Beznogov_2020}.}. Note that the information of the initial temperature distribution is rapidly lost, which means that the initial thermal state is irrelevant for long-term evolution studies \citep{Beznogov_2020}.

Boundary conditions at the origin and on the symmetry axis have mathematical motivation. They consist in requiring that $\widetilde{T}$ is regular and axisymmetric. Calculation of $\widetilde{T}$ in the limit $r \to 0$ requires that one integrates Eq.~\eqref{Eq:T_main} over a small spherical volume centered on $r=0$. 

At variance with this, surface boundary conditions have physical motivation. Following the standard approach in studies of NS cooling, we have chosen the boundary density $\rhob$ of the heat blanketing envelope (i.e., its bottom) to be  $10^{10}~\gcc$ \citep{PPP15}. This is the outer boundary where we impose the boundary condition (see Appendix~\ref{App:Envelope} for more details). 

The surface emission is determined by projecting the heat flux vector $h^i$, taken at the outer boundary, onto the normal to this boundary.

From Eqs.~\eqref{Eq:flux_red-shift}, \eqref{Eq:flux_r_def} and \eqref{Eq:flux_th_def} we obtain:
\begin{align}
	&\frac{\partial \widetilde{T}}{\partial r} = -\frac{A}{\mathrm{K}} \frac{N}{\Gamma} h^r, \\
	&\frac{\partial \widetilde{T}}{\partial \theta} = -\frac{A}{\mathrm{K}} \frac{N}{\Gamma} r h^\theta.
\end{align}
Now we need to define the unit normal vector (we are only interested in $r$ and $\theta$ \emph{covariant} components):
\begin{align}
	&n_r = A \left[ 1 + \frac{1}{R^2} \left( \frac{\dd R}{\dd \theta} \right)^2 \right]^{-\half}, 
	\label{Eq:norm_r}
	\\
	&n_\theta = -A \left[ 1 + \frac{1}{R^2} \left( \frac{\dd R}{\dd \theta} \right)^2 \right]^{-\half} \frac{1}{R} \frac{\dd R}{\dd \theta},
	\label{Eq:norm_th}
\end{align}
where $R=R(\theta)$ defines the radial coordinate of the outer boundary of the star, which by definition coincides with the bottom of the heat blanketing envelope (see Appendix~\ref{App:Envelope} for details). Since the heat flux is a contravariant vector, we can directly compute the scalar product:
\begin{align}
	\begin{split}
		h_\mathrm{s} = \vec{n} \cdot \vec{h} = &\mathrm{K} \frac{\Gamma}{N}
		\left[ 1 + \frac{1}{R^2} \left( \frac{\dd R}{\dd \theta} \right)^2 \right]^{-\half} \times\\
		&\left[
		-\left. \frac{\partial \widetilde{T}}{\partial r} \right|_{r=R} +
		\left. \frac{\partial \widetilde{T}}{\partial \theta} \right|_{r=R} \frac{1}{R^2} \frac{\dd R}{\dd \theta}
		\right].
	\end{split}
	\label{Eq:surf_flux}
\end{align}
On the other hand, the surface flux $h_\mathrm{s}$ can be computed via the heat-blanketing solution. In the plane-parallel approximation at any angle $\theta$\,\footnote{One should differentiate between local and redshifted quantities (the latter are denoted by a tilde).},
\begin{align}
	h_\mathrm{s}(\theta) = \sigma_\mathrm{SB} T^4_{\mathrm{s}}(\theta), \quad T_{\mathrm{s}} (\theta) = T_{\mathrm{s}}\left(T_{\mathrm{b}}(\theta) \right),
	\label{Eq:photon_flux}
\end{align}
where $\Ts \left(\Tb \right)$ is the ``$\Ts-\Tb$'' relation. For the latter we employ the envelope model of \cite{PCY97} with the parameter $\eta$ set to $10^{-18}$. This is equivalent to an iron (non-accreted) heat blanketing envelope. Combining Eqs.~\eqref{Eq:surf_flux} and \eqref{Eq:photon_flux} one obtains the surface boundary condition.

We have developed a code based on this formalism and called it \texttt{NSCool 2D Rot} to emphasize inheritance from \texttt{NSCool}, a publicly available NSs thermal evolution code in 1D  \citep{NSCool}. For more details, see Appendix~\ref{App:NSCool}.

\section{Models of rotating stars}
\label{sec:Models}

\subsection{EOS}
\label{ssec:EOS}

In order to identify the dependence on EOS, simulations are conducted for NSs built upon different models for the core EOS. For the sake of completeness, results corresponding to different models for the crust EOS will be commented as well.

\renewcommand{\arraystretch}{1.05}
\setlength{\tabcolsep}{5.0pt}
\begin{table*}
	\caption{Nuclear and astrophysical properties of our three EOS models. Listed are: binding energy per nucleon ($E_\mathrm{b}$); compression modulus ($K_{\mathrm{sat}}$); symmetry energy ($E_{\mathrm{sym}}$), its slope ($L_{\mathrm{sym}}$) and curvature ($K_{\mathrm{sym}}$) at the saturation point of symmetric matter ($n_{\mathrm{sat}}$). The remaining columns correspond to the maximum NS gravitational mass ($M_{\mathrm{G,max}}$); radius of $M_{\mathrm{G}} = 1.4~\Msun$ NS in \emph{Schwarzschild gauge}; range of combined tidal deformability for the GW170817 event; particle number density for the DURCA process threshold ($n_{DU}$) and the corresponding NS gravitational mass ($M_{\mathrm{DU}}$).}
	\centering
	\begin{tabular}{l c c c c c c c c c c c }
		\toprule
		Model & $E_\mathrm{b}$ & $n_{\mathrm{sat}}$ & $K_{\mathrm{sat}}$ & $E_{\mathrm{sym}}$ & $L_{\mathrm{sym}}$ & $K_{\mathrm{sym}}$ & $M_{\mathrm{G,max}}$ & $R_{14}$ & $\widetilde{\Lambda}$ & $n_{\mathrm{DU}}$ & $M_{\mathrm{DU}}$ \\
		      & (MeV) & $({\mathrm{fm}}^{-3})$ & (MeV) & (MeV) & (MeV) & (MeV) &  ($\Msun$) & (km) & & $({\mathrm{fm}}^{-3})$ & ($\Msun$)       \\
		\midrule
		APR               & $-16.00$ & 0.160 & 266.0 & 32.6 & 57.6 & $-102.6$ & 2.17 & 11.33 & 272--299   & 0.78 & 2.00 \\
		IUF               & $-16.40$ & 0.155 & 231.3 & 31.3 & 47.2 & $28.5$   & 1.95 & 12.64 & 499--530   & 0.61 & 1.77 \\
		NL3-$\omega \rho$ & $-16.24$ & 0.148 & 271.6 & 31.7 & 55.5 & $-8.1$     & 2.75 & 13.82 & 1042--1051 & 0.50 & 2.55  \\
		\hline
	\end{tabular}
	\label{Tab:EOS}
\end{table*}
\setlength{\tabcolsep}{6pt}
\renewcommand{\arraystretch}{1.0}

For the core we consider the following models: APR \citep{APR98}, IUF \citep{Fattoyev_2010} and NL3-$\omega \rho$ \citep{Horowitz_2001,Pais_2016}. Some of their characteristics, translated into values of symmetric saturated nuclear matter parameters and properties of cold catalyzed NSs, are presented in Table \ref{Tab:EOS}. Present uncertainties in the isoscalar and isovector channels, due to limited information from nuclear physics experiments, are accounted for by the ranges spanned by the parameters $K_{\mathrm{sat}}$ and $L_{\mathrm{sym}}$ and $K_{\mathrm{sym}}$, respectively. The models with the largest values of $K_{\mathrm{sat}}$ predict maximum NS masses much exceeding the lower limit on NS maximum masses, $\approx 2~\Msun$, while the model with the lowest value of $K_{\mathrm{sat}}$ is marginally consistent with this constraint. Radii of canonical mass NS, largely governed by the symmetry energy over $2 n_{\mathrm{sat}} \lesssim n_\mathrm{B} \lesssim 3 n_{\mathrm{sat}}$, span a 2.5 km interval.  The density dependence of the symmetry energy in the supra-saturation regime explains also the largely different density thresholds for the opening of DURCA processes. We nevertheless note that, due to the limited values of $L_{\mathrm{sym}}$, none of these EOSs allows for DURCA processes to operate in NSs with canonical masses \citep{Fortin_PRC_2016}. The less massive NS that allows for nucleonic DURCA process corresponds to the IUF EOS, i.e. the model with the largest value of $K_{\mathrm{sym}}$.  

Two EOSs are alternatively used for the crust. The first one corresponds to HZD \citep{Haensel_1989} supplemented by NV \citep{Negele_1973} models, covering the density domains below and above the neutron drip density, respectively. The second one corresponds to recent calculations by \citet{Mondal_PRC_2020} and \citet{Vinas_Sym_2021}; it is based on the finite range Gogny interaction D1M and accounts for shell and pairing corrections. These two EOS models are almost identical in what regards the $P(\rho)$ dependence but provide significantly different chemical compositions in the inner crust. More precisely, nuclei are more neutron rich in D1M than in HZD+NV, which entails a correspondingly lower density of unbound neutrons in the first model with respect to the second. Differences in the treatment of shell and pairing effects in addition to in-medium modification of cluster energy functionals are reflected also in different density dependencies of the atomic numbers though to a lesser extent. Core and crust EOSs are smoothly matched at $n_\mathrm{B}\approx n_\mathrm{sat}/2$. At this point we note that the calculation of fast rotating stars by LORENE requires the EOS to be sufficiently smooth. 

\subsection{Mechanical Structure}
\label{ssec:struct}

Before studying thermal evolution let us consider equilibrium configurations of rigidly rotating NSs. We choose to compare NSs with equal values of baryonic mass, a quantity that is conserved if accretion and mass loss are disregarded. Note that other choices, e.g. equal values of gravitational mass or central mass density, are also possible and have been used in the literature. 

The value of choice for the baryonic mass in this prospective study is $M_\mathrm{B} = 1.6~\Msun$, which results in gravitational masses close to the canonical value of $1.4~\Msun$. Two values of the rotation frequency are considered. The first one, 716 Hz, corresponds to the fastest-spinning millisecond pulsar known to date [PSR J1748--2446ad, see \citep{Hessels_2006}]. The second one corresponds to 99\% of the Kepler frequency and is EOS-dependent.

\renewcommand{\arraystretch}{1.05}
\setlength{\tabcolsep}{5.0pt}
\begin{table*}
	\caption{Properties of static and rigidly rotating NS with $M_\mathrm{B} = 1.6~\Msun$ built upon different EOS models.	Different rotation frequencies ($\Omega$) in Hz, specified in the second column, are assumed. The largest value corresponds, in each case, to 99\% of the Kepler frequency.
	Listed are: central density ($\rhoc$) in units of $10^{15}~\gcc$; 
	gravitational mass ($M_\mathrm{G}$) in units of ($\Msun$); 
	radius ($R$) and crust thicknesses ($\Delta R$) in km; 
	surface gravitational acceleration ($g_\mathrm{s14}$) in units of ($10^{14}$~cm~c$^{-2}$).
	Subscripts ``p'' and ``e'' stand for ``polar'' and ``equatorial'', respectively. Superscripts ``Co'' and ``Ph'' stand for ``coordinate'' and ``physical'' [radii and thicknesses]. Polar to equatorial axis ratio is denoted by $\zeta$ while $\zeta_\mathrm{cr}$ stands for polar to equatorial crust thickness ratio. As a reminder, quasi-isotropic coordinates differ from Schwarzschild ones. See text for details.}
	\centering
	\begin{tabular}{l r c c c c c c c c c c c}
		\toprule
		Model & $\Omega$ & $\rhoc$ &  $M_\mathrm{G}$   &  $R_\mathrm{p}^\mathrm{Co/Ph}$  & $R_\mathrm{e}^\mathrm{Co/Ph}$ & $\Delta R_\mathrm{p}^\mathrm{Co/Ph}$   &  $\Delta R_\mathrm{e}^\mathrm{Co/Ph}$  &  $\zeta^\mathrm{Co}$ & $\zeta_\mathrm{cr}^\mathrm{Co}$ & $g_{\mathrm{s14},\mathrm{p}}$ & $g_{\mathrm{s14},\mathrm{e}}$ \\
		\midrule
		APR               & 0    & 1.034 & 1.437 & 8.96/12.62  & 8.96/12.62  & 0.82/1.04 & 0.82/1.04 & 1.000 & 1.000 & 1.88 & 1.88 \\
		APR               & 716  & 0.989 & 1.446 & 8.46/11.93  & 9.57/13.39  & 0.80/1.02 & 1.02/1.28 & 0.884 & 0.785 & 1.92 & 1.44 \\
		APR               & 1096 & 0.890 & 1.465 & 7.61/10.69  & 11.85/16.11 & 0.80/1.03 & 2.21/2.69 & 0.642 & 0.363 & 1.92 & 0.26 \\
		IUF               & 0    & 0.817 & 1.452 & 10.15/13.80 & 10.15/13.80 & 0.94/1.16 & 0.94/1.16 & 1.000 & 1.000 & 1.50 & 1.50 \\
		IUF               & 716  & 0.740 & 1.463 & 9.37/12.72  & 11.27/15.13 & 0.92/1.14 & 1.31/1.60 & 0.831 & 0.700 & 1.53 & 0.98 \\
		IUF               & 932  & 0.657 & 1.478 & 8.61/11.62  & 13.48/17.69 & 0.93/1.17 & 2.49/2.97 & 0.639 & 0.375 & 1.51 & 0.20 \\
		NL3-$\omega \rho$ & 0    & 0.525 & 1.463 & 11.40/14.98 & 11.40/14.98 & 1.39/1.69 & 1.39/1.69 & 1.000 & 1.000 & 1.22 & 1.22 \\
		NL3-$\omega \rho$ & 716  & 0.489 & 1.478 & 10.05/13.23 & 13.14/17.06 & 1.34/1.64 & 2.19/2.61 & 0.765 & 0.612 & 1.27 & 0.62 \\
		NL3-$\omega \rho$ & 826  & 0.470 & 1.486 & 9.47/12.44  & 14.80/18.99 & 1.34/1.65 & 3.28/3.87 & 0.640 & 0.409 & 1.27 & 0.16 \\
		\hline
	\end{tabular}
	\label{Tab:Models}
\end{table*}
\setlength{\tabcolsep}{6pt}
\renewcommand{\arraystretch}{1.0}

Table \ref{Tab:Models} lists some of the properties of NSs. For completeness characteristics of static stars are provided as well. NS radii and crust thicknesses are reported along the equatorial (``e'') and polar (``p'') directions. Two values are provided. The first corresponds to the quasi-isotropic spherical coordinates used by \texttt{LORENE} (superscript ``Co''); the second corresponds to what a local observer would measure (superscript ``Ph'') and, thus, refers to proper (``physical'') lengths. Note that (i) coordinate lengths measured in quasi-isotropic spherical coordinates can not be compared to those derived in Schwarzschild coordinates, (ii) proper (physical) lengths are the same regardless of the choice of coordinates. Proper radial distances (i.e. with $t,~\theta,~\varphi =$~const) can be computed as
\begin{align}
	l_\mathrm{r} = \int_{r_\mathrm{min}}^{r_\mathrm{max}} A\left(r', \theta \right) \dd r'.
	\label{Eq:proper_length}
\end{align}
A word of caution is needed in what regards the radii and crust thicknesses mentioned in Table~\ref{Tab:Models}. Both are measured up until the bottom of the envelope, i.e., up until $\rhob=10^{10}~\gcc$. See Appendix~\ref{App:Envelope} for more details.

Let us first focus on static (i.e., non-rotating spherically symmetric) configurations. The following remarks are in order:
(i) the highest (lowest) central density value is obtained for the most (least) compact configuration, which in its turn corresponds to the softest (stiffest) EOS model; note that characterization of the EOS as soft or stiff in relation to $M_\mathrm{B} = 1.6~\Msun$ NS refers to the slope of the $P(\rho)$ curve over the density domain span in these stars and, in general, differs from what is obtained when much wider domains are considered, as it is the case of the maximum mass configuration (see Table~\ref{Tab:EOS});
(ii) as already discussed in relation to Table \ref{Tab:EOS} uncertainties in the density dependence of the symmetry energy in the supra-saturation domain lead to substantial variations in the ranges of predicted radii;
(iii) the scatter between the values of gravitational mass is very limited, less than $4\%$;
(iv) even if the same EOS model is used for the crust, its thickness changes from one model to another;
(v) the EOS model that provides the largest (smallest) value for the radius produces the thickest (thinnest) crust.

\begin{figure*}
	\includegraphics[]{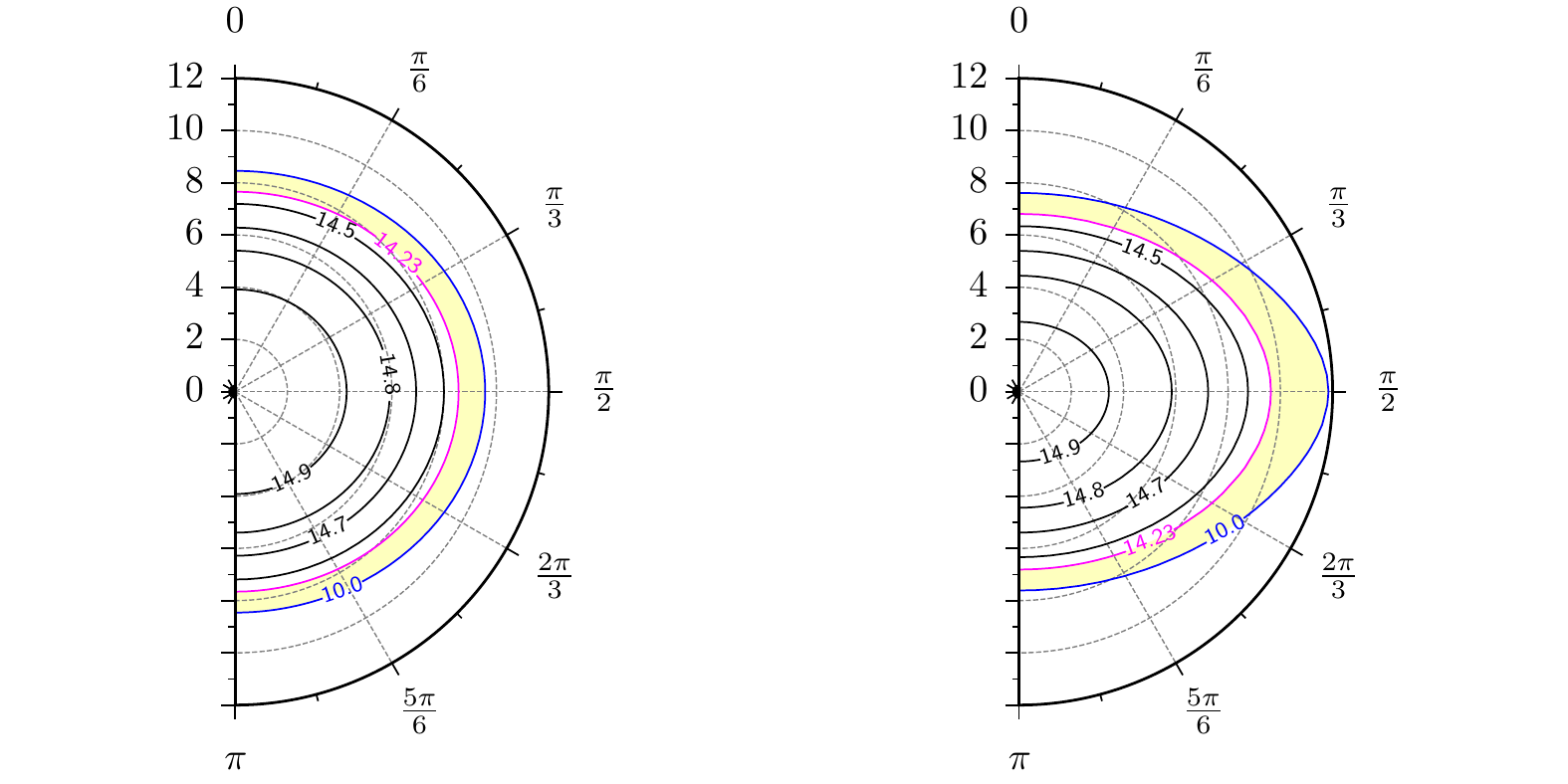}
	\caption{Mechanical structure of rotating NSs built upon the APR EOS. Left (right) panel: $\Omega = 716$ Hz (1096 Hz) model. The contour lines show the density distribution as a function of coordinate radius $r$ (in km) and polar angle $\theta$ (in radians). Values on the contours indicate the logarithm of density. The light yellow shadowed area indicates the crust. The crust-core boundary is depicted by a magenta line ``14.23'' and blue contour ``10'' depicts the bottom of the heat blanketing envelope which is the outer boundary of the star.}
	\label{Fig:Rho_2D_APR}
\end{figure*}

Let us now consider the case of the mass shedding limit. The largest (smallest) frequency ($\Omega_\mathrm{K}$) corresponds to the APR (NL3-$\omega\rho$) EOS. These are the EOSs with the most (least) compact static configuration. The dispersion of $\Omega_\mathrm{K}$ is significant while that of the polar to equatorial axis ratio ($\zeta$) is insignificant. Indeed, the Kepler frequency of the APR EOS model exceeds that of the NL3-$\omega\rho$ EOS model by 25\%, while the difference between the axis ratios is just ~0.3\%. One notes that the APR and NL3-$\omega\rho$ EOSs provide for both static and maximally rotating stars the smallest and largest values of radii, respectively. Central densities of maximally rotating configurations are lower by 10\% (NL3-$\omega\rho$ EOS) to 20\% (IUF EOS) than those of the static configurations and no clear correlation between $\Omega_\mathrm{K}$ and $\rhoc$ can be established. Being dilute the crust is much more affected by rotation than the core: at the equator it increases by 130\% (NL3-$\omega\rho$ EOS) to 160\% (APR EOS); the crust at the pole is much thinner than the crust at the equator.

NSs rotating with 716~Hz feature properties intermediate between those of static and maximally rotating stars. As in non-rotating and maximally rotating NSs, the largest (smallest) value of central density corresponds to the APR (NL3-$\omega\rho$) EOS and the most (least) deformed configuration is obtained for the NL3-$\omega\rho$ (APR) EOS. The dispersion among the values of $\zeta$ in NSs built upon different EOSs is rather important at 716~Hz, while practically no dispersion is obtained when approaching Kepler frequency. Quite interestingly, the ratios between crust thickness at the pole and at the equator, $\zeta_{\mathrm{cr}}$, produced by various EOSs are in a different relation than their counterparts at the Kepler limit. 

\begin{figure}
	\includegraphics[]{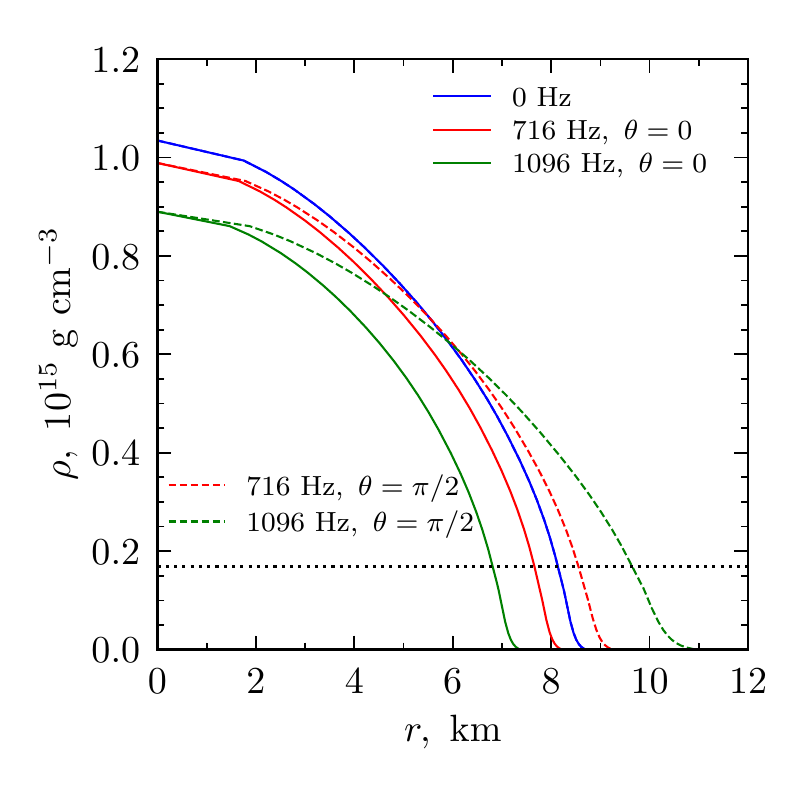}
	\caption{Radial profiles of density along the polar (solid lines) and equatorial (dashed lines) directions for a static NS (blue line) as well as NSs rotating at 716~Hz (red lines) and 1096~Hz (green lines). The considered EOS model is APR. The horizontal dotted line marks the crust-core transition density.}
	\label{Fig:Rho-r_APR}
\end{figure}

Further insight into the geometrical properties of rotating NSs and matter distribution is shown in Figs.~\ref{Fig:Rho_2D_APR} and \ref{Fig:Rho-r_APR}. The considered EOS model is APR; the rotation frequencies are mentioned in the key legend and figure caption. Contour lines in Fig.~\ref{Fig:Rho_2D_APR} show the spatial distribution of matter density; values on the plot correspond to the logarithm of density. The yellow shadowed area indicates the crust; it extends from the crust-core boundary (magenta line marked ``14.23'') to the bottom of the heat blanketing envelope which coincides with the outer boundary of the star (blue contour ``10''). The star rotating at 716~Hz (1096~Hz) deviates little (significantly) from spherical symmetry. Consequently, crust thickness is marginally (sizably) larger at the equator than at the pole. All these features are in agreement with the data in Table~\ref{Tab:Models}. We anticipate that these differences in the crust thickness will translate into different heat diffusion timescales that will modify the cooling pattern in the equatorial direction compared to the polar direction.

Fig.~\ref{Fig:Rho-r_APR} first confirms what we have already seen in Table~\ref{Tab:Models}. Namely, that rapid rotation makes NSs less compact. It also shows that density gradients along the equatorial direction are smaller than those along the polar direction. 

Before leaving this section some comments are in order. First, we assume that $\Omega =$~const throughout the cooling process, which is equivalent to discarding any spin-down, e.g.  due to magnetic braking \citep{Negreiros_2017}. It is a reasonable simplification as magnetic fields in isolated NSs are not necessarily strong. The second comment deals with the highest rotation frequency we have considered.

For a NS rotating at near-Kepler frequency any small deviation from axisymmetric configuration immediately leads to rapid loss of angular momentum due to GWs emission. These deviations can be caused by various reasons: r-mode instability, f-mode-instability, thermal or magnetic mountains, etc. Both theoretical computations and the fact that there are currently no known pulsars rotating faster than 716~HZ, suggest that the spin-down due to GWs emission is extremely efficient and that NSs rotating at the near-Kepler limit cannot maintain their rotation frequency long-term. Thus, studying long-term cooling of such stars is academic. We do not expect to find any. Still, we included them in our analysis as a limiting case opposite to the limiting case of non-rotating spherically symmetric stars. We want to check to the maximum possible extent the effects caused by rotation.

Mechanical properties of NSs built upon the D1M EOS of the crust are identical to those obtained when the HDZ+NV EOS is used. The reason is that the $P(\rho)$ dependence of both EOSs is virtually the same. Note nevertheless that differences in chemical composition lead to slight differences in the thermal evolution, see Sect.~\ref{sec:Cooling}.

\section{Cooling of rotating isolated NSs}
\label{sec:Cooling}

We shall consider here the standard long-term cooling of purely nucleonic isolated NSs with $M_\mathrm{B} = 1.6~\Msun$ built upon the EOS models described in Sec.~\ref{ssec:EOS}. The results presented in this section are obtained using the HZD+NV EOS for the crust. We have also performed cooling simulations with the D1M crust EOS and we shall briefly comment on them at the end of this section. As mentioned in Sect.~\ref{sec:Intro}, we solve mechanical structure equations only once for each model (using \texttt{LORENE}, see Sect.~\ref{ssec:metric}) and then solve only thermal evolution equations at each time step (using \texttt{NSCool 2D Rot}, see Appendix~\ref{App:NSCool}) [keeping the structure fixed throughout the evolution].

Moderate values of $L_{\mathrm{sym}}$ and NS mass ensure that none of the considered NS models allows for the DURCA process. Nucleonic pairing in the crust and core is disregarded for simplicity. This means that neutrino cooling by the formation and breaking of Cooper pairs does not occur either. So, the main driving force behind the thermal evolution is the modified Urca process and, at later stages, surface photon emission. The neutrino emission rates, thermal conductivities and specific heat capacity are standard and described by, e.g., \cite{Yakovlev_1980,YKGH01,Shternin_2007,HPY07}; for implementation see \citep{PLPS04,PPLS11}. Magnetic fields and heat sources are disregarded. 

\begin{figure*}
	\includegraphics[]{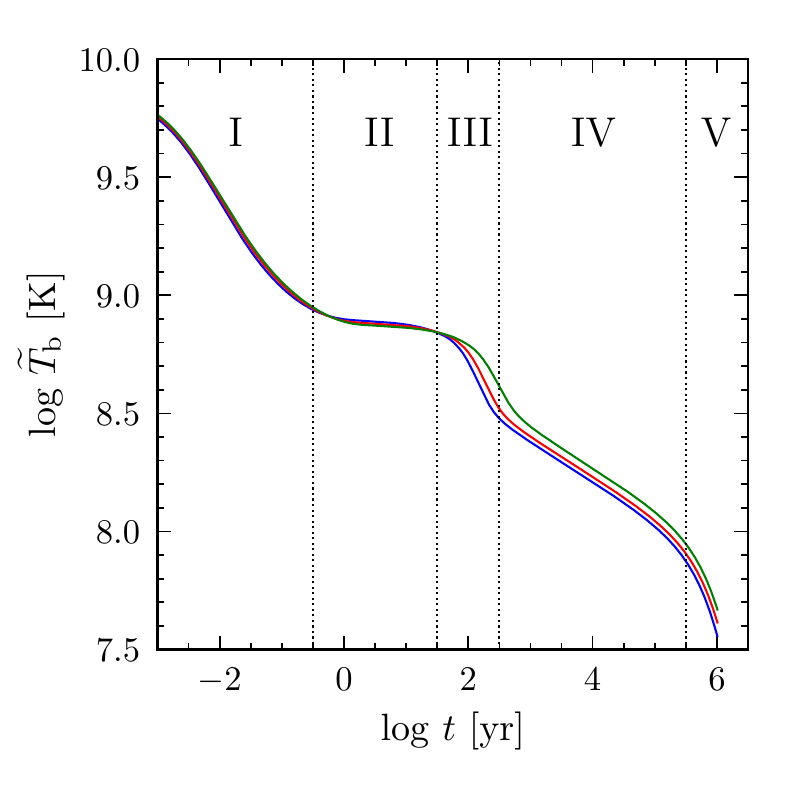}
	\includegraphics[]{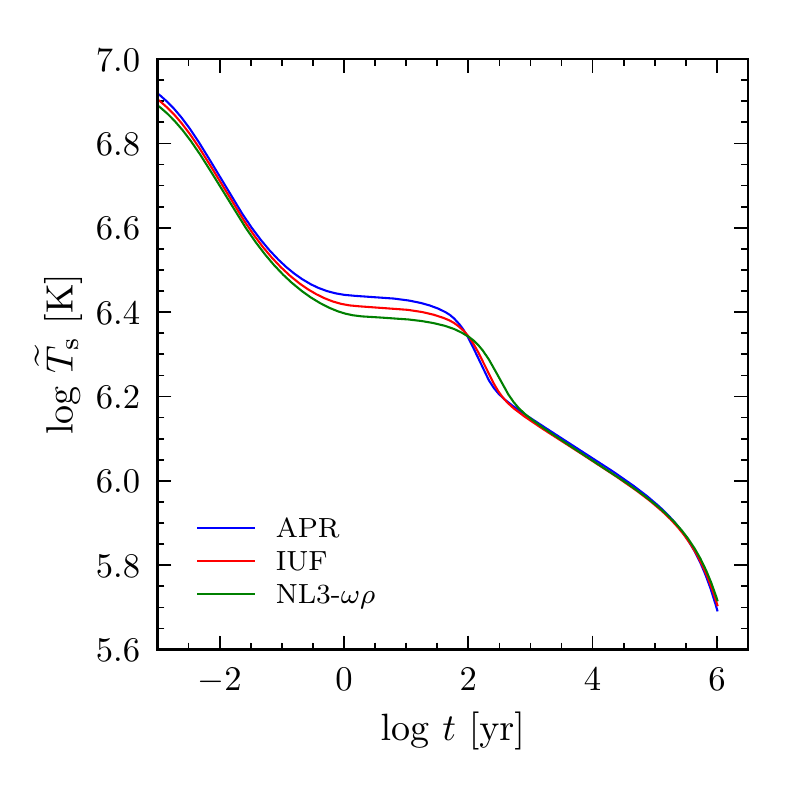}
	\caption{Redshifted temperature at the bottom of the envelope (left panel) and on the surface (right panel) as a function of time (cooling curves). Results correspond to static NSs with $M_\mathrm{B} = 1.6~\Msun$ built upon the EOS models indicated in the legend.}
	\label{Fig:Cool_curves_NoRot}
\end{figure*}

To start with let us consider the limiting case of static NSs. Redshifted temperatures at the bottom of the envelope, $\widetilde{T}_\mathrm{b}$, and on the surface, $\widetilde{T}_\mathrm{s}$, are plotted as a function of time in Fig.~\ref{Fig:Cool_curves_NoRot}, left and right panels, for the different EOS models. The left panel shows that during the phases of relaxation from initial conditions\,\footnote{We will refer to everything happening before the early plateau phase as relaxation from initial conditions. Note however, that the actual relaxation happens much faster and is followed by some other (sub-)phases before the early plateau phase begins. See \citep{Beznogov_2020} for details.}, (stage~I with $\log t \left[ {\mathrm{yr}} \right] \lesssim -0.5$), and early plateau (stage~II with $-0.5 \lesssim \log t\left[ {\mathrm{yr}} \right] \lesssim 1.5$), the various stars have almost identical values of the bottom temperature. The crust relaxation phase lasts for the longest (shortest) period for the NS built upon the NL3-$\omega\rho$ (APR) EOS. The similarity between $\widetilde{T}_\mathrm{b}$ can be explained by the identical composition of the crusts, while the duration of the crust relaxation phase reflects the crust thickness. Indeed, as shown in Table \ref{Tab:Models}, static stars built upon the NL3-$\omega\rho$ and APR EOSs have the thickest and thinnest crusts, respectively, which means that the cold wave from the core needs less time to pass through the crust in the case of the NS with the APR EOS than in the one of the NS with the NL3-$\omega\rho$ EOS. The model that is the first (last) one to leave the early plateau phase is characterized by the lowest (highest) $\widetilde{T}_\mathrm{b}$ during the subsequent crust-core thermalization (stage~III with $1.5 \lesssim \log t\left[ {\mathrm{yr}} \right] \lesssim 2.5$) and neutrino cooling (stage~IV for $t\left[ \mathrm{yr} \right] \gtrsim 2.5$) stages.

Comparing the left and right panels, one can judge the impact of the heat blanketing envelope solution that is used as an outer boundary condition in the cooling calculations. $\widetilde{T}_\mathrm{s} (t)$ produced by our NS models are identical in stages I and IV; in stage II the highest (lowest) values of $\widetilde{T}_\mathrm{s}$ correspond to APR (NL3-$\omega\rho$) EOS; in stage III the ranking is inverted. The re-arrangement of cooling curves is due to the dependence of the envelope solution on surface gravitational acceleration, see Table~\ref{Tab:Models}. 

\begin{figure*}
	\includegraphics[]{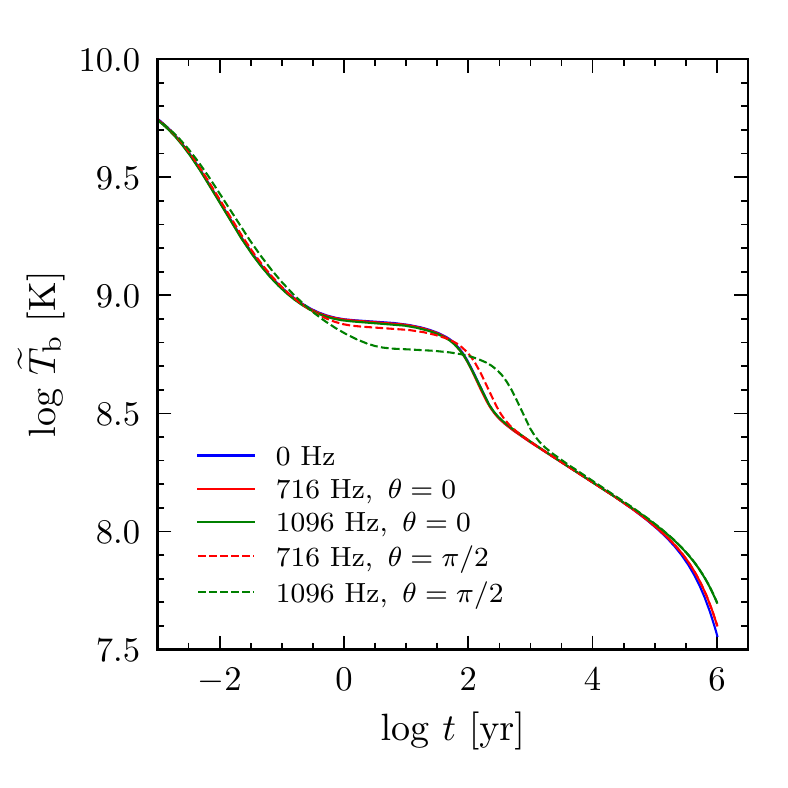}
	\includegraphics[]{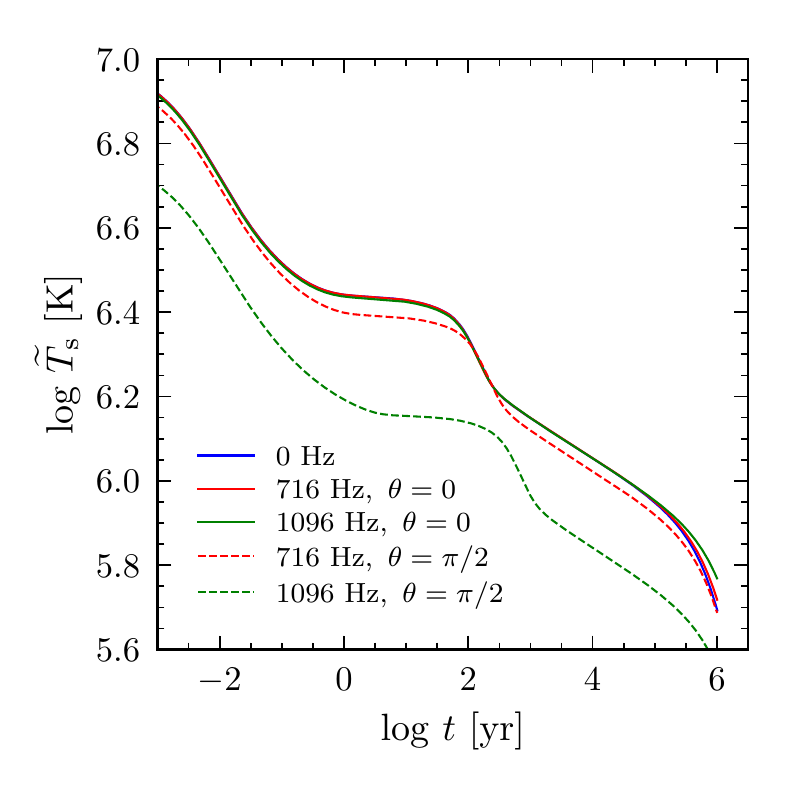}
	\caption{Redshifted temperature at the bottom of the envelope (left panel) and on the surface (right panel) as a function of time (cooling curves). Results correspond to the APR EOS and different rotation frequencies specified in the legend. Considered are values at the pole ($\theta=0$) and at the equator ($\theta=\pi/2$).}
	\label{Fig:Cool_curves_APR}
\end{figure*} 

The modification of $\widetilde{T}_\mathrm{b/s} (t)$ under rotation is investigated in  Fig.~\ref{Fig:Cool_curves_APR} for the particular case of the APR EOS. Redshifted values of the temperature at the bottom of the envelope (left panel) and on the surface (right panel), corresponding to $\Omega=716~{\mathrm{Hz}}$ and 1096 Hz, are confronted at $\theta=0$ (pole) and $\pi/2$ (equator). Also considered is the case of a static star. From the left panel it turns out that $\widetilde{T}_\mathrm{b,p}(t)$ is affected by rotation only in the photon cooling stage (V) and for extreme rotation frequencies. This is not very surprising as the polar regions are less affected by rotation and heat transfer there is mostly radial (unlike equatorial regions, see below). At variance with this, a series of modifications affect $\widetilde{T}_\mathrm{b,e}(t)$: the redshifted value of the temperature at the bottom of the envelope is slightly enhanced in stage I; stage II is delayed and lasts longer; $\widetilde{T}_\mathrm{b,e}^{\mathrm{(II)}} < \widetilde{T}_\mathrm{b,p}^{\mathrm{(II)}}$, $\widetilde{T}_\mathrm{b,e}^{\mathrm{(III)}} > \widetilde{T}_\mathrm{b,p}^{\mathrm{(III)}}$, stage IV is delayed. The increase in the duration of stage II is related to the increased thickness of the crust in equatorial regions: a thicker crust can store more energy and also has a larger heat diffusion time (which goes as the square of the length-scale). As a result, the cold wave from the core needs more time to punch through the crust. This explanation was previously put forward by \citet{Miralles_1993, Schaab_1998}.

The right panel of Fig.~\ref{Fig:Cool_curves_APR} demonstrates that gravitational acceleration on the surface is accountable for much stronger modifications of $\widetilde{T}_\mathrm{s,e}$ under rotation. At variance with this, no rotation induced effect is visible on $\widetilde{T}_\mathrm{s,p}$. Overall and over the whole time span redshifted temperature at the equator is lower than the one at the pole and the difference between the cooling curves increases with $\Omega$.

A word of caution is in order here. For non-rotating NSs the redshifted surface temperature is the one that --- except effects related to atmosphere structure, interstellar absorption, etc. --- a distant observer would see. Some authors assume that this is also the case for rotating NSs. This is however not obvious. Indeed, \cite{Vincent_2018} demonstrated that the observed spectra depend on the line of sight and that the spectral distortion, i.e. deviation from color-corrected blackbody spectrum, is small for uniform surface temperature distribution. This obviously does not apply for a fast rotating NS. A rigorous approach to study this question would be to perform general-relativistic ray tracing, similar to that performed in \cite{Vincent_2018}, but this is beyond the scope of the current study.

\renewcommand{\arraystretch}{0.0}
\setlength{\tabcolsep}{0.0pt}
\begin{figure*}
	\refstepcounter{figure}
	\begin{tabular}{c c}
		& \multirow{3}{*}{\rotatebox[origin=l]{90}{\parbox{\textheight}{\textbf{Figure~\thefigure}. Redshifted temperature distributions in rotating NSs built upon the APR EOS. Left (right) panels correspond to $\Omega= 716$~Hz (1096~Hz). The radial coordinate is translated into the logarithm of density. Top, middle and bottom panels correspond to the ages of 10, 150 and 300~yr, respectively.}}} \\
		\includegraphics[]{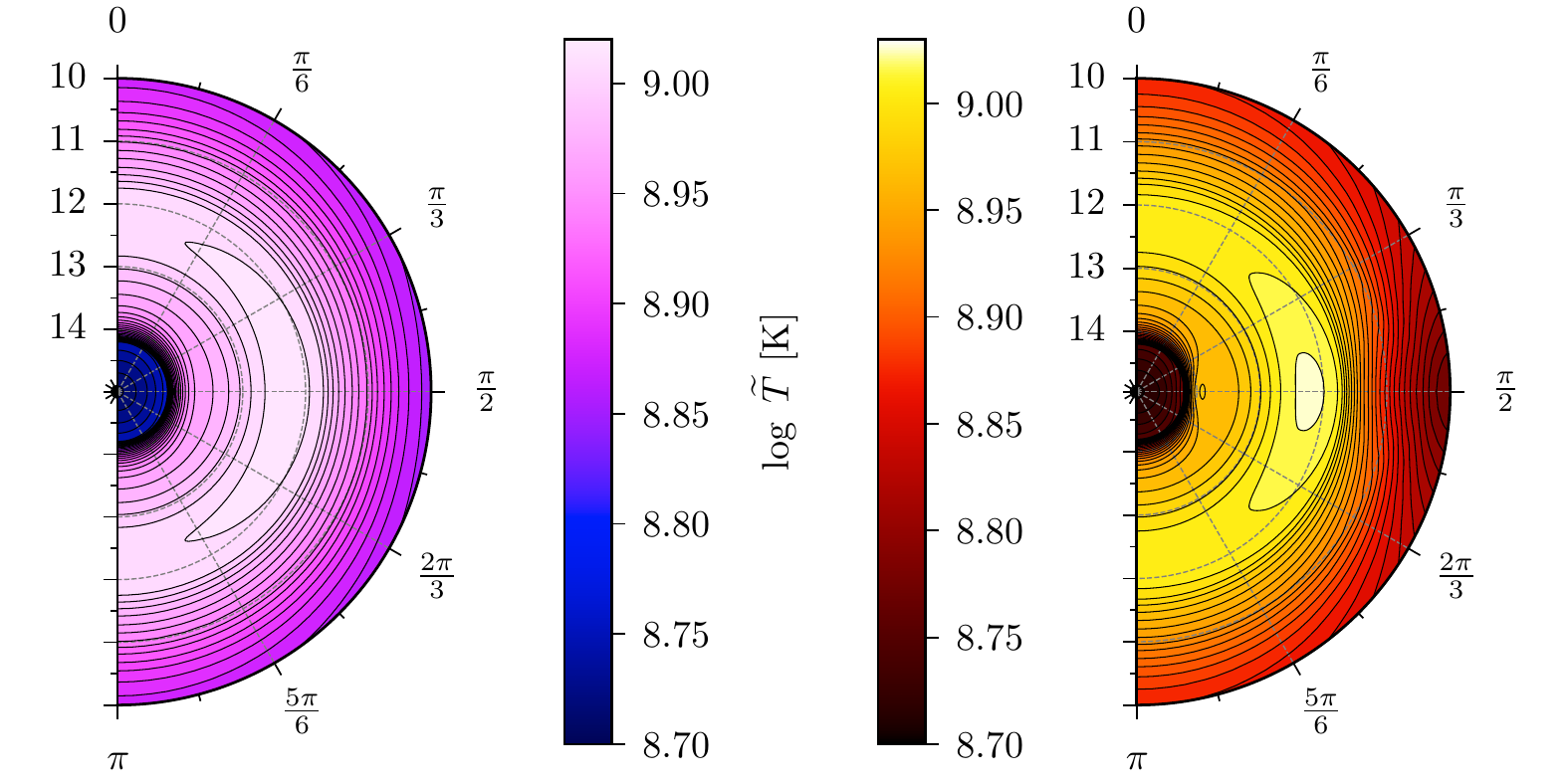}  & \\ 
		\includegraphics[]{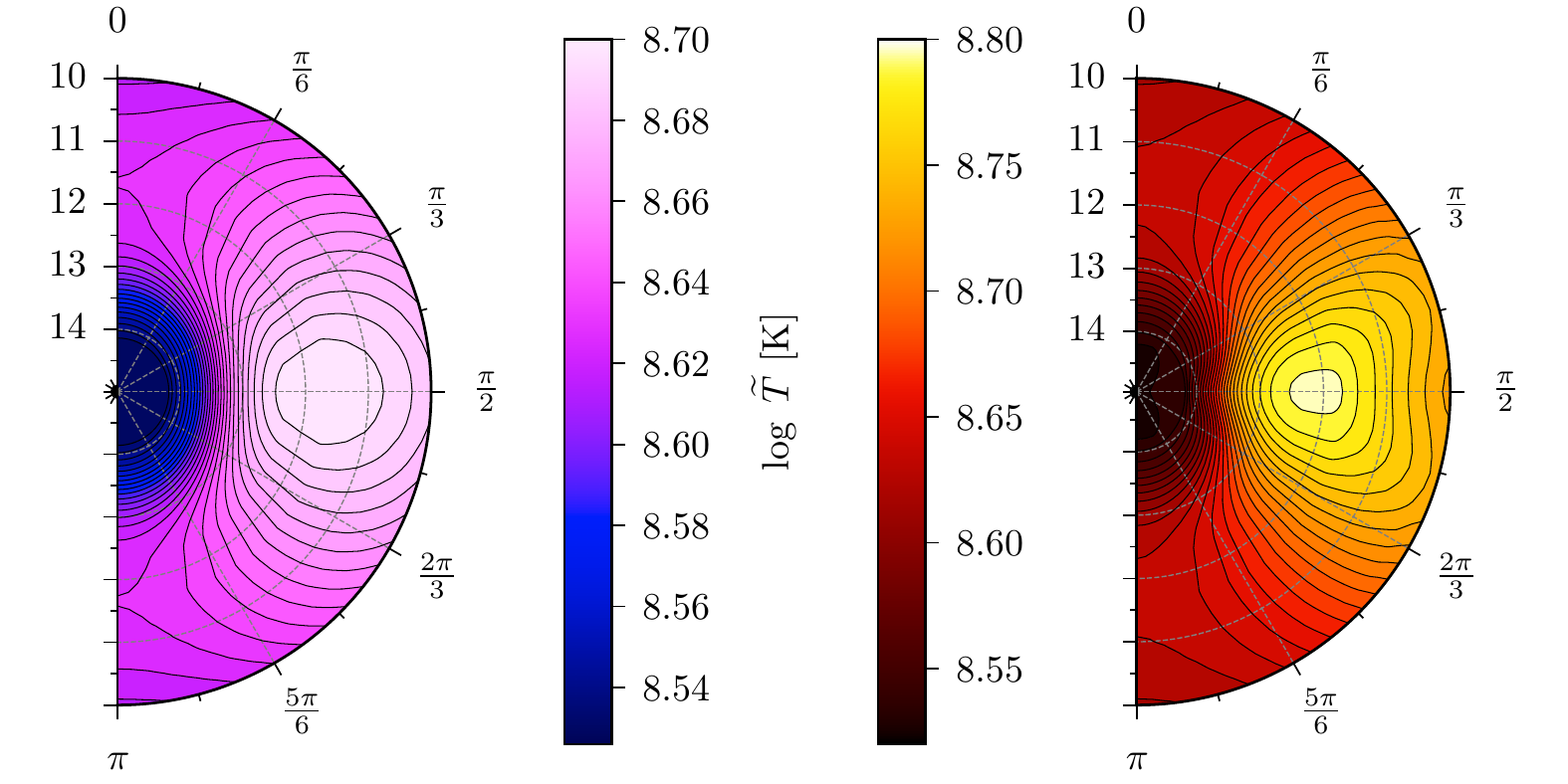} & \\
		\includegraphics[]{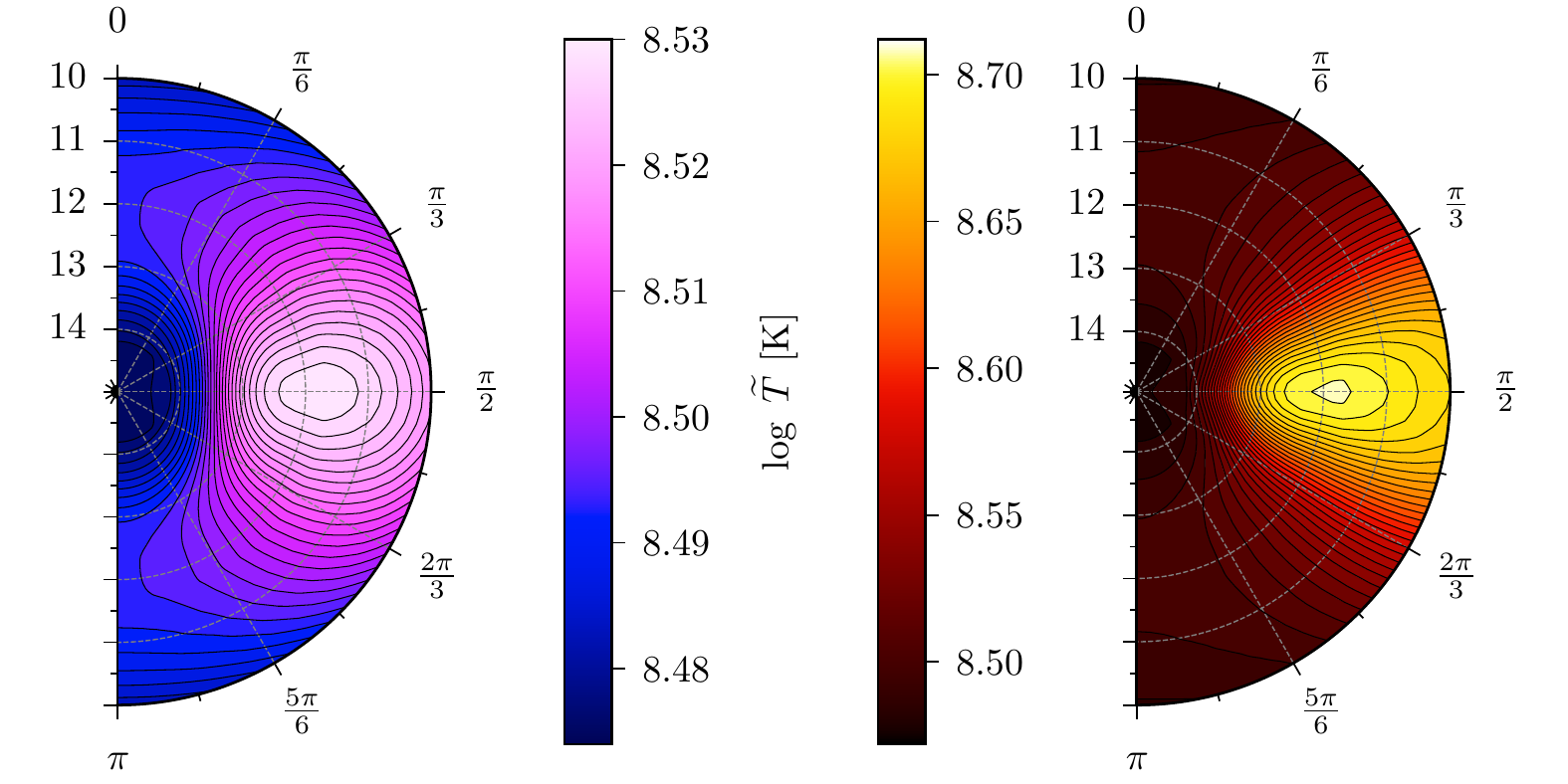} & \\
	\end{tabular}			
	\label{Fig:T_2D_APR}
\end{figure*}
\renewcommand{\arraystretch}{1.0}
\setlength{\tabcolsep}{6.0pt}
\begin{figure*}
	\includegraphics[]{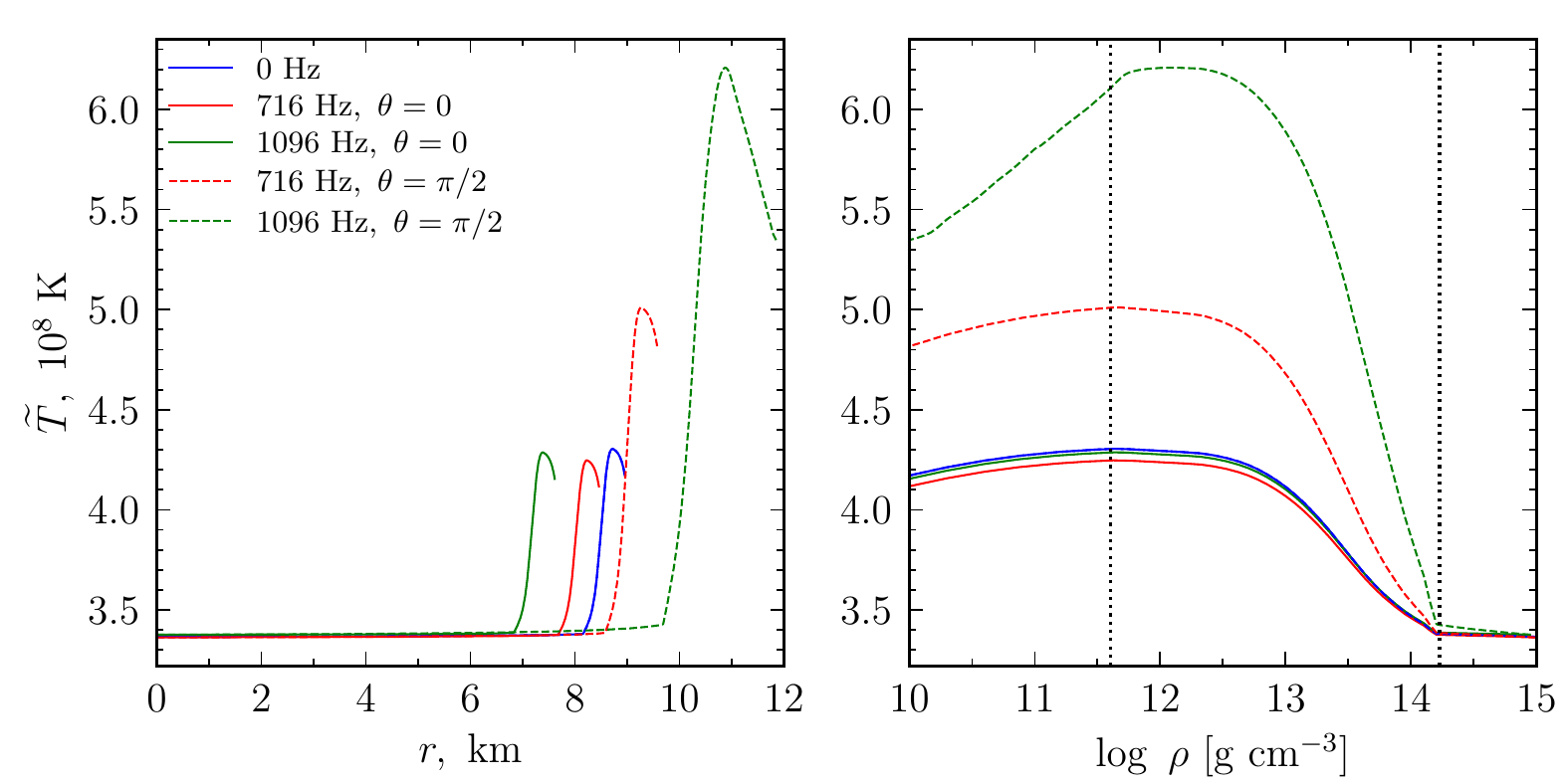}
	\caption{Profiles of redshifted temperature in a $M_\mathrm{B} = 1.6~\Msun$ NS built upon the APR EOS. Results corresponding to different rotation frequencies, indicated in the legend, are confronted at the pole and at the equator at $t = 150$~yr. Representations as a function of coordinate radius (logarithm of density) are provided in the left (right) panel. The two vertical dotted lines in the right panel indicate the neutron drip density and the crust-core transition density, respectively.}
	\label{Fig:T-r-rho_APR_t=150}
\end{figure*}
\begin{figure*}
	\includegraphics[]{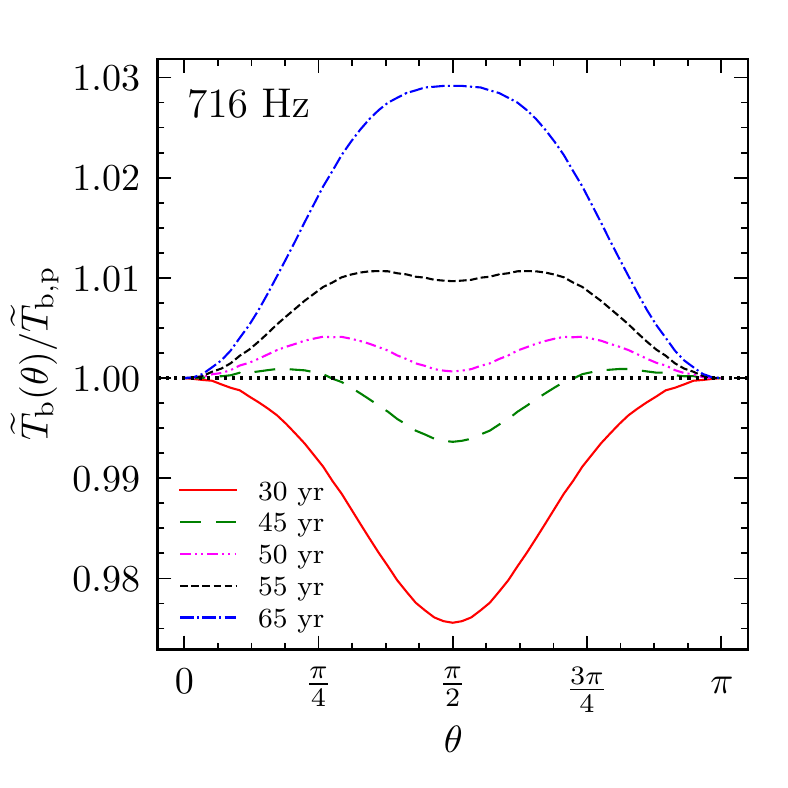}
	\includegraphics[]{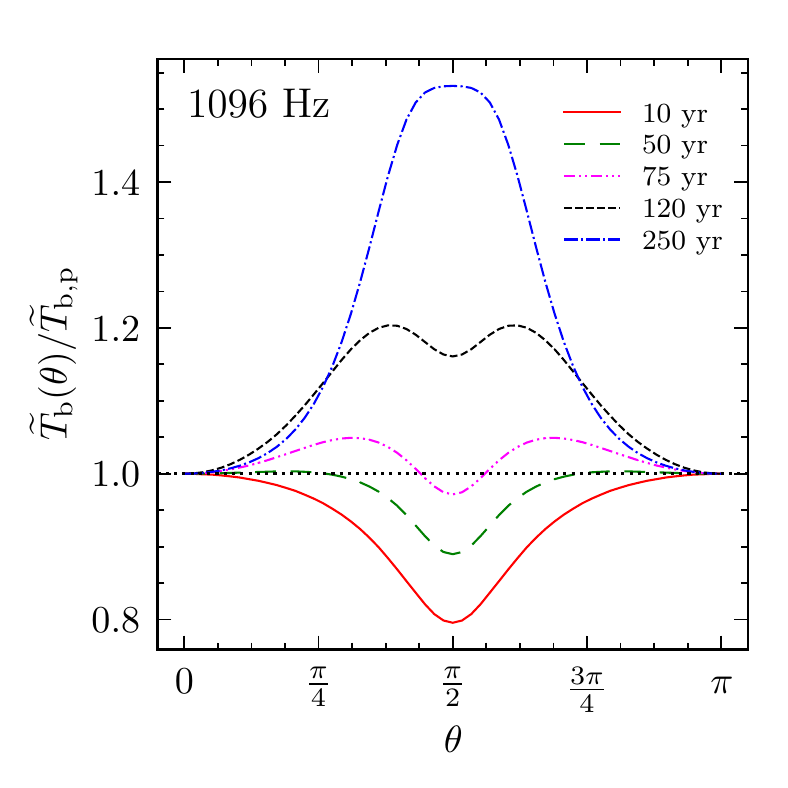}
	\caption{Redshifted temperature at the bottom of the envelope as a function of polar angle $\theta$ at different instants, as shown in the legends. Results corresponding to the $M_\mathrm{B} = 1.6~\Msun$ NS built upon the APR EOS rotating at 716~Hz (1096~Hz) are illustrated in the left (right) panel. Note that the values on the Y axis are normalized to the corresponding values at the pole (``p'' subscript).}
	\label{Fig:Tb-theta_APR}
\end{figure*}

A detailed look at the internal redshifted temperature distribution is offered in Fig.~\ref{Fig:T_2D_APR}. The considered stars are built upon the APR EOS; they rotate at $\Omega = 716$~Hz (left column) and $\Omega = 1096$~Hz (right column). Notice that the radial coordinate is the logarithm of density, which hinders information on star deformation. Different instants in the cooling evolution are considered: $t=10$~yr (top), $t=150$~yr (middle) and $t=300$~yr (bottom panels). 

One can see that at the age of 10~yr (stage II) the temperature gradient is mostly radial and only slightly affected by the rotation. At the age of 150~yr (stage III) the situation is very different. In the core and innermost crust (densities around $10^{14}~\gcc$) the temperature gradient is still mostly radial while at lower densities ($10^{10} \lesssim \rho \lesssim 10^{12}~\gcc$) in the equatorial zone the temperature gradient is mostly angular. A ``heat blob'' appears at the equator. It acts almost as a ``point source'' of heat that flows in all directions from it. For the faster rotating star the effect is more pronounced and the temperature distribution more complex. At the age of 300~yr (stage III) the situation is qualitatively similar to the one at 150~yr. However, for the slow rotating star the temperature distribution is already almost uniform, as one would expect from a star that enters the neutrino cooling phase; for the fast rotating star the effects of the ``heat blob'' are still present in the equatorial regions producing both radial and angular temperature gradients.

Despite complicated temperature distributions, the main idea behind the ``heat blob'' is rather simple. During the phases I--III thermal evolution of the crust is governed both by its heat conduction and neutrino emission. Rotation causes the crust to be thicker in the equatorial regions. Thus, the total heat capacity of the crust in this region increases due to the increase in its volume. Total neutrino losses increase for the same reason. Yet the heat exchange can only increase proportionally to the increase in the area of interfaces between the equatorial region and the rest of the star. Obviously, the increase in volume dominates over that of the area. Thus, the equatorial region becomes a reservoir of heat simply because it can store more thermal energy and cannot transfer it to the rest of the star fast enough (heat conduction is limited by the area).

\begin{figure*}
	\includegraphics[]{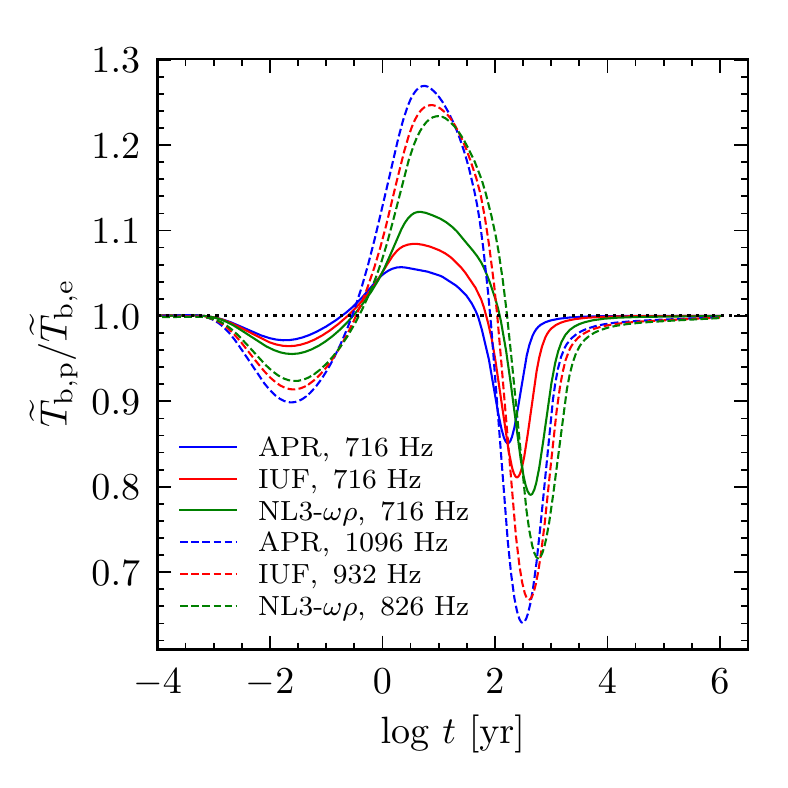}
	\includegraphics[]{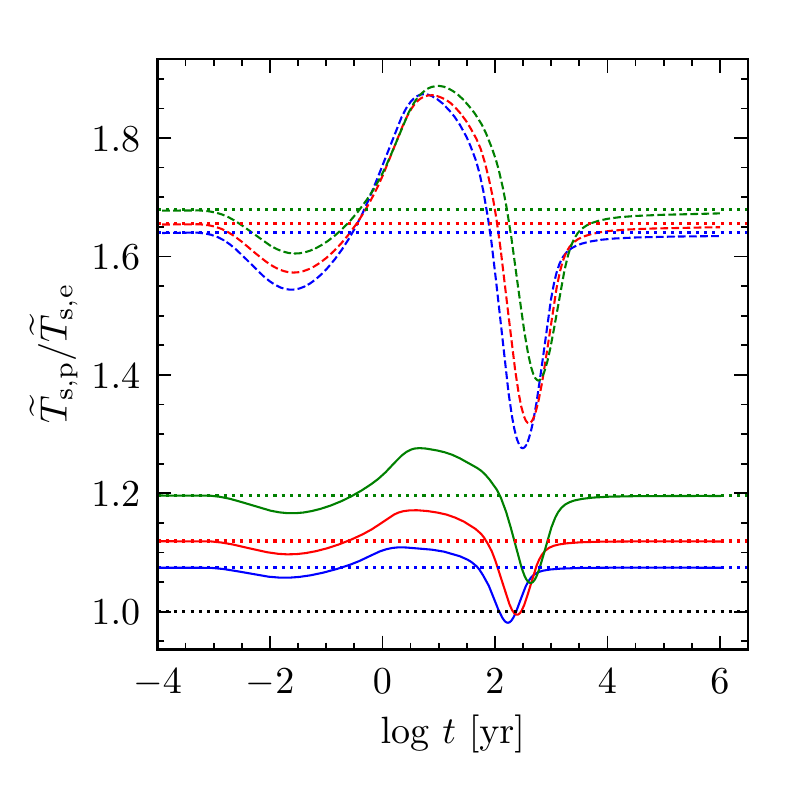}
	\caption{Ratio of polar to equatorial redshifted temperature as functions of time for NS with $M_\mathrm{B} = 1.6~\Msun$ rotating with different frequencies that are indicated in the legend.
		Left (right) panels correspond to the bottom of the envelope (surface). Various EOS models, introduced in Table \ref{Tab:EOS}, are employed. The horizontal black dotted line in the left panel marks the unity. The remaining horizontal dotted lines in the right panel correspond to the ratio $(g_{\mathrm{s},\mathrm{p}}/g_{\mathrm{s},\mathrm{e}})^{\slfrac{1}{4}}$; their color legend is the same as for EOS models.}
	\label{Fig:T_p+e-t_All} 
\end{figure*}

Fig.~\ref{Fig:T-r-rho_APR_t=150} further investigates the distribution of redshifted temperature in rotating stars built upon the APR EOS at $t=$150~yr, previously considered in the middle panels of Fig. \ref{Fig:T_2D_APR}. Results along the polar and equatorial directions for $\Omega=716~{\mathrm{Hz}}$ and 1096~Hz are confronted with those of static stars. Representations as functions of radial coordinate (density) are shown in the left (right) panel. The core is already isothermal: there is no radial temperature gradient and the values at $\theta=0$ and $\theta=\pi/2$ coincide. The situation at the crust is the opposite: a non-monotonic radial profile is obtained and $\widetilde{T}_\mathrm{p} < \widetilde{T}_\mathrm{e}$.  The temperature along the polar direction depends only loosely on the rotation frequency while a strong rotation frequency dependence manifests along the equatorial direction. In the left panel the displacement on the $X$-axis between curves corresponding to different directions and rotation frequencies is evocative of NS deformation. The fact that the temperature along the equator is higher than the one at the pole is obviously due to the ``heat blob''. But, as we shall see later, even higher redshifted temperatures are obtained at certain instants over $0<\theta < \pi/2$. Their values as well as those of $\theta$ obviously depend on the EOS, rotation frequency and time. The general feature is that the dispersion in $\widetilde T(r, \theta, t)$ increases with the rotation frequency. 

The angular distribution of the redshifted temperature at the bottom of the envelope and different instants is depicted in Fig.~\ref{Fig:Tb-theta_APR}; the same rotating stars as in Fig.~\ref{Fig:T-r-rho_APR_t=150} are considered. The instants for which the curves are plotted are different in the left and right panels; they correspond either to the early plateau (stage~II) or crust-core thermalization (stage~III) stages and show the same patterns. For early and mid-stage~II the temperature at the equator is lower than at the pole. As soon as the stars enter the stage~III the situation becomes inverted. At intermediate instants $\widetilde{T}_{\mathrm{b}}(\theta)$ shows a complex structure and the highest temperature is obtained for $0<\theta<\pi/2$. The excursion of $\widetilde{T}_{\mathrm{b}}(\theta)$ augments with the rotation frequency; the time the NS needs to enter the neutrino cooling phase (stage~IV) does the same. All these features agree with what we have seen in Fig.~\ref{Fig:Cool_curves_APR}. 

The ratio between redshifted temperatures\,\footnote{The ratio of local temperatures is exactly the same as the redshift factor $N/\Gamma$ [see Eq.~\eqref{Eq:T_red-shift}] does not depend on $\theta$ along the iso-contours of density (for barotropic EOSs this corresponds to the first integral of mechanical structure, derived from the conservation of energy-momentum).} at the pole and at the equator is represented as a function of time in Fig.~\ref{Fig:T_p+e-t_All} (cf. also Fig. 2 of \citealt{Schaab_1998}). A wide time span is considered. Left (right) panels depict values at the bottom of the envelope (on the surface). In addition to the NS built upon APR EOS, we have also plotted the curves for other EOSs here. Similarly to the previous figures results at 716 Hz and 99\% of the Kepler frequency are depicted. Both panels confirm the complex behavior of redshifted temperature as a function of time, rotation frequency and polar angle previously seen in Fig.~\ref{Fig:Tb-theta_APR}, which has considered a much narrower time span. They also show a significant EOS dependence. Again, in agreement with Fig.~\ref{Fig:Tb-theta_APR} the amplitude of the variation of $\widetilde{T}_{\mathrm{b/s,p}}/\widetilde{T}_{\mathrm{b/s,e}}$ increases with the rotation frequency. All curves manifest two minima and one maximum; the instants when these are obtained are the same for the quantities represented in the left and right panels. The explanation relies on the fact that the ``$\Ts-\Tb$'' relation is instantaneous. The two minima are obtained for $\log t [\mathrm{yr}] \approx -1.5$ and $2 \lesssim \log t [\mathrm{yr}] \lesssim 3$ corresponding to phase~I and phase~III, respectively. They are characterized by $\widetilde{T}_{\mathrm{b,p}}/\widetilde{T}_{\mathrm{b,e}} < 1$. The maximum occurs over $0 \lesssim \log t [\mathrm{yr}] \lesssim 1$ and is characterized by $\widetilde{T}_{\mathrm{b,p}}/\widetilde{T}_{\mathrm{b,e}} > 1$. Early ($\log~t[\mathrm{yr}] \lesssim -3$) and late ($\log~t[\mathrm{yr}] \gtrsim 4$) in the thermal evolution $\widetilde{T}_{\mathrm{b}}$ is the same at the pole and at the equator. At early moments this reflects that the initial conditions are the same, see Sect.~\ref{ssec:boundcond}. The situation at late moments is due to the fact that the star has become isothermal.

The fact that, in the crust-core thermalization phase, $\widetilde{T}_{\mathrm{b,p}} < \widetilde{T}_{\mathrm{b,e}}$ has been previously noted by \citet{Miralles_1993} and \citet{Schaab_1998}, who explained it in terms of the time needed for the cold wave traveling from the core to reach the crust and then punch through it. We remind that the core cools faster than the crust because it emits more neutrinos; in rotating stars the crust is thicker at the equator than anywhere else, see Table~\ref{Tab:Models}, Figs~\ref{Fig:Rho_2D_APR} and \ref{Fig:Rho-r_APR}. 

Concerning the EOS dependence we note that, around the extrema, the ordering (in amplitude) of $\widetilde{T}_{\mathrm{b,p}}/\widetilde{T}_{\mathrm{b,e}}$ at 716~Hz is opposite to the one at 99\% of Kepler frequency. Analysis of data in Table \ref{Tab:Models} reveals that the values of this ratio are correlated with $\zeta_{\mathrm{cr}}$. This finding highlights once more the role of the crust in the thermal evolution. We also note that, at 716~Hz and the near Kepler limit, the maximum at $0 \lesssim \log t [\mathrm{yr}] \lesssim 1$ and the minimum at $2 \lesssim \log t [\mathrm{yr}] \lesssim 3$ are reached at the earliest (latest) times in NS built upon the APR (NL3-$\omega\rho$) EOS, for which the crust thickness is the smallest (largest), Table \ref{Tab:Models}. 

\looseness=-1
A comparison of the left and right panels allows one to see the influence of the heat blanketing envelope and its dependence on the surface gravitational acceleration \citep{Beznogov_2021}. For better visibility the ratio between surface gravitational acceleration at the pole and at the equator, to the power $\slfrac{1}{4}$, is represented with dotted horizontal lines in the right panel; the same color convention as for $\widetilde{T}_{\mathrm{s,p}}/\widetilde{T}_\mathrm{s,e}$ is used. $\left(g_\mathrm{s,p}/g_\mathrm{s,e}\right)^{1/4}$ depends on the oblateness of the star and, thus, on EOS stiffness; for correlation among them, see Table~\ref{Tab:Models} and the comments in Sect.~\ref{ssec:struct}. The right panel in Fig. \ref{Fig:T_p+e-t_All} shows that, in most of the considered circumstances (i.e. EOS models and rotation frequencies) the redshifted surface temperature at the pole is higher than its counterpart at the equator; though, as shown by the APR and IUF curves at 716~Hz over $2 \lesssim \log t [\mathrm{yr}] \lesssim 2.5$ exceptions from this rule can occur.   

\begin{figure}
	\includegraphics[]{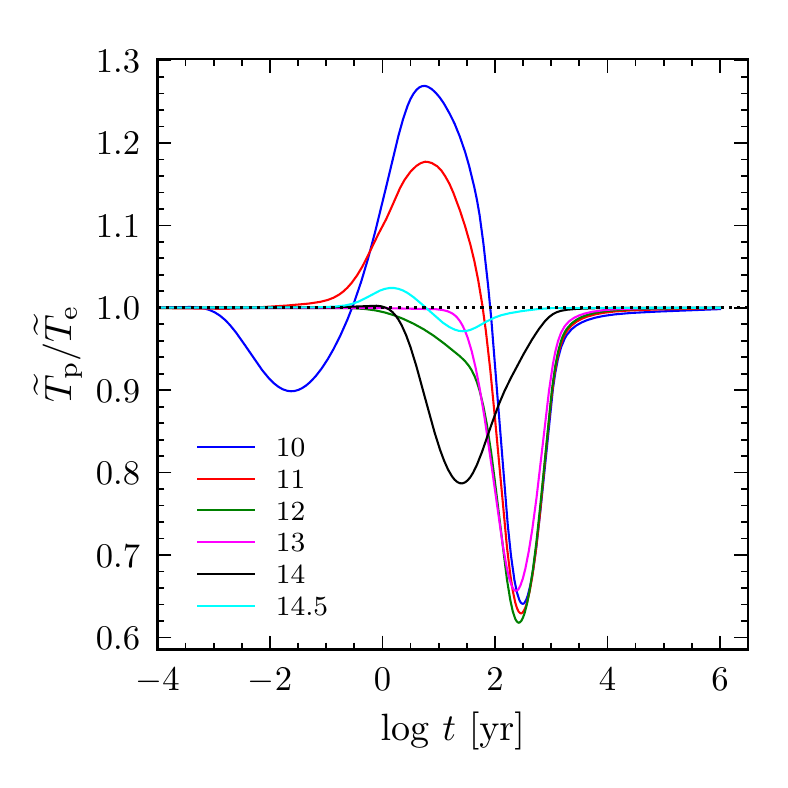}
	\caption{The ratio of polar to equatorial redshifted temperature as a function of time at different densities inside the NS. The considered EOS model is APR. The star rotates at 1096~Hz. The considered values of density are indicated in the legend (in $\log$). The curve marked ``14.5'' corresponds to the core while all the others correspond to the crust.}
	\label{Fig:T_p+e-t_APR_depth}
\end{figure}

\looseness=-1
Information complementary to that shown in Fig.~\ref{Fig:T_p+e-t_All} (left panel) is provided in Fig.~\ref{Fig:T_p+e-t_APR_depth}, where the ratio between redshifted temperatures at the pole and at the equator is reported, as a function of time, for different densities inside the crust as well as one density in the core (labeled ``14.5''). Only the case of the NS built upon APR and rotating at the (near) Kepler limit is considered. We note that the core is almost isothermal; the largest deviation from constant redshifted temperature is obtained at the bottom of the envelope; the lower the density the more complex the temperature profile; high-density shells reach thermal equilibrium earlier than low-density shells. The minimum at $ \log t[\mathrm{yr}] \approx -1.5$ is obtained only for $\rho=10^{10}~\gcc$. Thus, it is likely some sort of reaction to the surface boundary condition. Taking into account that the minimum appears at the age of $t \ll 1$~yr we speculate that it might be not physical. Indeed, the heat diffusion timescale through a heat blanketing envelope with a bottom density of $\rhob = 10^{10}~\gcc$ is a few months to a year (see, e.g., \citealt{Beznogov_2021}) and variations of temperature happening on a shorter timescale are not reliable in such model. 

Before leaving this section let us comment on the sensitivity to the composition of the inner crust. A comparison of the cooling curves obtained using the D1M and HZD+NV EOSs for the crust indicates that only the crust-core thermalization phase (phase III) is affected. More precisely, our results show that when D1M is used, phase III starts by $\approx 20$~yr earlier.

%

\section{Conclusions}
\label{sec:Concl}

We have investigated the standard cooling of isolated nucleonic NSs uniformly rotating with frequencies up to the mass shedding limit. The EOS dependence of thermal evolution was studied by considering different EOS models.

Our results indicate that due to rotation-induced deformation a ``heat blob'' is formed in the crust and persists around the equatorial plane over $1 \lesssim \log t [\mathrm{yr}] \lesssim 3$, which roughly corresponds to phases II-III. Its exact location and extension depend on the NS deformation, composition and temperature profiles which, in turn, are determined by the core and crust EOSs, rotation frequency and time in thermal evolution. As a consequence the redshifted temperature manifests a complex distribution throughout the interior as well as at the bottom of the envelope. In particular, our results indicate that -- depending on time -- the maximum of $\widetilde T_{b}$ can be obtained at the pole, at the equator or for intermediate values of polar angle. Large scale structures in the temperature distribution can be explained in terms of the above-mentioned ``heat blob'' and delays in cooling phases in equatorial regions compared to polar regions and static stars can be attributed to heat diffusion timescales. Both the ``heat blob'' and heat diffusion timescales depend on the specific heat, thermal conductivity and geometry of the crust. 

In contrast with the crust, the core is little affected by rotation. Indeed, even for near-Kepler models the polar to the equatorial relative difference in temperature does not exceed a few percent. This situation can be explained by a combination of different factors:
(i) being more compact, the core is less deformed;
(ii) the core's heat conductivity is much higher than the crust's one, which means that the core requires less time in order to become isothermal;
(iii) the temperature evolution in the core is driven primarily by neutrino emission from its whole volume, which is not affected by rotation. 

The redshifted surface temperature $\widetilde{T}_\mathrm{s}(\theta)$, despite being determined by the redshifted bottom temperature $\widetilde{T}_\mathrm{b}$, might still show a different angular pattern. The explanation is that the surface gravity acceleration, which depends on the polar angle, enters the ``$\Ts-\Tb$'' relation.

Our simple explanation in terms of the ``heat blob'' is unable to account for some finer details of the temperature distribution and evolution for near-Kepler models. Yet this is not a critical issue as studying long-term cooling of near-Kepler models without taking the spin-down into account is a  rather academic exercise. It is still useful as it allowed us to set limiting cases of the possible effects of rotation on long-term cooling.

Microphysics and observational issues not addressed in the present work, as nucleonic pairing in the core and in the crust and the relation between observed and redshifted surface temperatures, will be considered elsewhere.

\begin{acknowledgments}
The authors thank D. Yakovlev for useful discussions. MB and AR acknowledge support from a grant of the Ministry of Research, Innovation and Digitization, CNCS/CCCDI -- UEFISCDI, Project No. PN-III-P4-ID-PCE-2020-0293, within PNCDI III. DP and MB acknowledge financial support by the Mexican Consejo Nacional de Ciencia y Tecnolog\'ia with a CB-2014-1 grant \#240512 and the Universidad Nacional Aut\'onoma de M\'exico through an UNAM-PAPIIT grant \#IN109520. MB also acknowledges support from a postdoctoral fellowship from UNAM-DGAPA. JN gratefully acknowledges the Italian Istituto Nazionale di Fisica Nucleare (INFN), the French Centre National de la Recherche Scientifique (CNRS), and the Netherlands Organisation for Scientific Research for the construction and operation of the Virgo detector and the creation and support of the EGO consortium.
\end{acknowledgments}

\software{
	\begin{itemize}[nosep]
		\item  \texttt{NSCool 2D Rot} [not publicly available]
		\item  \texttt{LORENE} [\citealt{LORENE_2016}, \url{https://lorene.obspm.fr}]
		\item  \texttt{Python} [\url{https://www.python.org}] with packages:
		\begin{itemize}[nosep]
			\item  \texttt{matplotlib} [\cite{Matplotlib}, \url{https://matplotlib.org}] 
			\item  \texttt{f90wrap} [\cite{f90wrap}, \url{https://github.com/jameskermode/f90wrap}]
			\item \texttt{numpy} [\cite{NumPy}, \url{https://numpy.org}]
			\item  \texttt{scipy} [\cite{SciPy}, \url{https://scipy.org}]
		\end{itemize}
	\end{itemize}
}

\appendix
\section{Heat blanketing envelope \& long-term thermal evolution of NSs}
\label{App:Envelope}

In this paper we have employed a widely accepted approach to handle long-term thermal evolution of NSs. The main idea was described in Sect.~\ref{sec:Intro} and here we want to briefly address some finer points regarding the heat blanketing envelope and surface boundary condition. 

\looseness=-1
As mentioned in Sect.~\ref{sec:Intro}, the very outer layers of a NS are not degenerate enough and their mechanical structure depends noticeably on temperature. A conventional solution to circumvent this issue is to split the star into two domains: the star itself and a heat blanketing envelope. This split significantly simplifies all calculations. The idea dates back to $\sim 1980$-s to pioneering works by \cite{GPE82,GPE83}. The envelope goes from the actual surface of the star until its bottom at some density $\rhob$, which is also called boundary density. The particular choice of $\rhob$ depends on the specifics of the task at hand, like the required heat diffusion timescale through the envelope (see below), the presence of magnetic fields, etc. We adopted  the ``standard'' value of $\rhob = 10^{10}~\gcc$. Other choices are possible, but beyond the scope of the discussion here. 

The envelope is thin, contains $\sim 10^{-7}~\Msun$ of matter and has a negligible impact on the mechanical structure of the star. However, unlike the rest of the star, the thickness of the envelope depends strongly on the temperature and varies from $\sim 10$~m in cold stars to a few hundred meters in very hot stars with near-Eddington luminosity \citep{Beznogov_2021,Beznogov_2020}. Thus, to have a temperature-independent notion of the star's radius it is rather common to measure it from the center until $\rhob$. The same reasoning applies to crust thickness. Since the envelope is thin, the change in radius is marginal; crust thickness is somewhat more affected. The values of radii and crust thicknesses presented in Table~\ref{Tab:Models} are calculated this way. 

While the envelope has a negligible impact on the mechanical structure of the star, it plays an important role in thermal evolution. Thus, the question of heat transport through the envelope should be addressed.  Under the conditions discussed in, e.g., \cite{GPE83}, which are usually satisfied for long-term cooling, heat transport through the envelope can be computed \emph{separately} in quasi-stationary plane-parallel approximation employing a temperature-dependent EOS of Coulomb plasma \citep{PC10}. The result of such \emph{separate} calculation is the relation between the temperature at the top of the envelope and the temperature at its bottom ($\Tb$). Now, by definition, the temperature at the top is the surface temperature ($\Ts$). Thus, we have obtained the ``$\Ts - \Tb$'' relation. Two comments are necessary here as they were mentioned in Sect.~\ref{sec:Cooling}:
(i) the functional dependence $\Ts(\Tb)$  [or its inverse $\Tb(\Ts)$] is, of course, instantaneous. One should keep in mind, however, that in reality heat needs some time to diffuse through the envelope. For $\rhob = 10^{10}~\gcc$ the heat diffusion timescale is $\sim 1$~yr. Thus, one cannot study temperature variations occurring on a timescale faster than this; 
(ii) as shown in \cite{GPE82,GPE83} to a very high accuracy $\Tb$ is not actually a function of $\Ts$, but rather of $\Ts/g_\mathrm{s}^{\slfrac{1}{4}}$ ($g_\mathrm{s}$ being the surface gravity acceleration). This is very convenient as it allows calculating ``$\Ts - \Tb$'' relation for some value of $g_\mathrm{s}$ and then rescaling it for any other value. We have used this feature extensively, as for rotating star $g_\mathrm{s}$ varies with polar angle $\theta$.
       
Now, from the point of view of the cooling computation, the outer boundary of the star is the bottom of the envelope, i.e. the outer boundary density is $\rhob = 10^{10}~\gcc$. This is where we define the surface boundary condition of the main cooling calculation. And this is another reason to measure radii and crust thicknesses up until the outer boundary of the main calculation. To relate this outer boundary to the actual surface one employs the ``$\Ts - \Tb$'' relation (see surface boundary condition equations in Sect.~\ref{ssec:boundcond}).  

The last question to consider here is whether $\rhob$ is high enough that thermal effects on the mechanical structure can be ignored at densities $\rho \ge \rhob$. The answer is straightforward: Fermi temperature of electron gas at $\rho = \rhob = 10^{10}~\gcc$ is $T_\mathrm{F} \approx 10^{11}$~K (for ions with mass to charge numbers ratio equal to 2). The contribution of thermal effects is of the order $(\pi T/T_\mathrm{F})^2$ \citep{Baym_1981}. Thus, even at the very beginning of the cooling calculation with the initial temperature at the bottom of the envelope set to $10^{10}$~K, this contribution is less than 10\%. After a few more days the temperature drops more than twice and thermal contribution becomes totally negligible. Obviously, a few days are completely irrelevant for cooling on the timescales of years or more.

For more details on the envelope and how the ``$\Ts - \Tb$'' relations are calculated, see the pioneering works by \cite{GPE82,GPE83} and/or the recent review by \cite{Beznogov_2021} and references therein.

\section{NSCool 2D Rot}
\label{App:NSCool}

\texttt{NSCool 2D Rot} is a NSs thermal evolution code developed by two of us (MB and DP). It is a major upgrade of one-dimensional thermal evolution code \texttt{NSCool} by \cite{NSCool}. Our code is written fully in FORTRAN 90. All functions and subroutines ``inherited'' from \texttt{NSCool} were re-written in FORTRAN 90. Our code makes extensive use of FORTRAN 90 derived types and employs parallelization via OpenMP in the main solver.

The numerical scheme is based on a finite difference approach. The spatial discretization uses a second-order central differences scheme with physical quantities (e.g., temperature, density, etc.) defined in the grid points, first spatial derivatives defined in the middle between the grid points and second spatial derivatives\,\footnote{Since thermal conductivity is assumed to be scalar in the present work, there are no mixed second derivatives as can be explicitly seen from Eqs.~\eqref{Eq:T_main}, \eqref{Eq:Grr} and \eqref{Eq:Gthth}.} defined once again in the grid points; this approach is similar to the staggered grids used in computational fluid dynamics. Unlike other existing 2D NSs thermal evaluation codes, the grid in \texttt{NSCool 2D Rot} explicitly includes points on the symmetry (i.e., rotation) axis as well as the grid point in the star center.  

The temporal discretization is done with a fully implicit backward Euler scheme. Despite being only first order (in time), this method is L-stable [see, e.g., \cite{Hairer_1996}], which is crucial for solving the very stiff and highly nonlinear Eq.~\eqref{Eq:T_main}. We have tested trapezoidal and midpoint (semi-)implicit A-stable second-order schemes and found that they produce small oscillations in the solution during photon cooling phase [which was not totally unexpected, see, e.g., \cite{Bui_1979}]. While some second or higher-order L-stable methods do exist [see, e.g., \cite{Bui_1979,Bulatov_2011}], their implementation is \emph{significantly} more involved than for the backward Euler scheme. Taking into account that \texttt{NSCool} also relies on backward Euler temporal discretization and has proven to have no issues with the accuracy, we decided that this method is sufficient for our purposes. Nevertheless, we have performed accuracy tests by comparing calculations done with different timesteps.

Nonlinear algebraic equations obtained after spatial and temporal discretization are handled by our custom-written solver. It employs the Newton-Raphson method with a full Jacobian, which is computed by numerical differentiation via a second-order central difference scheme and taking into account its structure (i.e., only non-zero elements are computed and stored). The linearized equations at each Newton iteration are handled by our custom-written modification of generalized minimal residual solver [GMRES, \cite{Saad_1986}] with the tridiagonal part of the Jacobian as the preconditioner. This rather complicated setup with nested iterations (outer Newton iterations and inner GMRES iterations) was chosen for two main reasons: 
(i) the size of the Jacobian makes direct linear solvers too slow and its complicated structure prevents the employment of specialized banded or sparse matrix solvers [due to integration needed for the boundary condition at the origin, nine-diagonal structure of the Jacobian is supplemented by a column and a row]
(ii) other solution strategies like simple fixed-point iterations [see, e.g., \cite{Hoffman_2001}] or the alternating-directions implicit method [ADI, see, e.g., \cite{Peaceman_1955}] were either not converging or converging very slowly during our initial tests.

Before using \texttt{NSCool 2D Rot} for practical applications, we have performed extensive tests of the solver itself, checked  the overall robustness of the spatial and temporal discretization and compared the results with \texttt{NSCool} for spherically symmetric configurations.

To check the solver we exported the function that generates a set of discretized nonlinear algebraic equations at each timestep to \texttt{Python} using \texttt{f90wrap} tool. Then we solved exported equations employing \texttt{scipy.optimize.root} at exactly the same timesteps as in \texttt{NSCool 2D Rot} and compared the resulting 2D temperature distributions. We set the tolerances (maximum $\delta \widetilde{T}/\widetilde{T}$, where $\delta \widetilde{T}$ is a Newton-Raphson correction to  $\widetilde{T}$) identically in both solvers (i.e., in \texttt{Python} and \texttt{NSCool 2D Rot}) and equal to $10^{-10}$. The maximum relative difference in $\widetilde{T}$ between two solvers solving exactly the same equations did not exceed $3 \times 10^{-10}$ for all timesteps before the age of $\sim 50$~yr (further in time \texttt{Python} solver was not converging).

Then, we compared cooling curves computed by \texttt{NSCool 2D Rot} with different spatial grids and timesteps. Doubling the number of grid points in both the $r$ and $\theta$ directions (i.e., 4 times more grid points in total) resulted in a maximum relative change in $\widetilde{T}_{\mathrm{s,e}}$ of $\sim 0.8\%$. This is on a par with the behavior of \texttt{NSCool} where doubling the number of grid points leads to  an $\sim 0.7\%$ maximum relative change in surface temperature. Doubling the number of timesteps resulted in an $\sim 1.6\%$ maximum relative change in $\widetilde{T}_{\mathrm{s,e}}$. Again, this is on a par with \texttt{NSCool} and was expected as the requested temperature change between consecutive timesteps\,\footnote{This ``user defined'' parameter is a part of an adaptive timestep selection algorithm.} was $5\%$ and $2.5\%$. So it is not surprising that the two calculations differ from each other by a comparable amount.

Finally, we have compared cooling curves produced by \texttt{NSCool 2D Rot} and \texttt{NSCool} for identical spherically symmetric systems. The maximum relative difference in $\widetilde{T}_{\mathrm{s}}$ was $\sim 6.4\%$. On the scale of, e.g., Fig.~\ref{Fig:Cool_curves_NoRot} such difference would not be seen (the curves would be on top of each other). For any practical applications this difference is irrelevant. Overall, we consider that the performance of the solver is excellent and employed spatial and temporal discretization scheme is on a par with \texttt{NSCool}.

The current implementation of \texttt{NSCool 2D Rot} was designed to work together with \texttt{LORENE}, but this is not mandatory as the codes do not communicate with each other directly. The results of \texttt{LORENE} (metric functions, matter distribution, etc.) are simply stored in files that are read by \texttt{NSCool 2D Rot}.

\bibliographystyle{aasjournal}
\bibliography{Cool_Rot}
\end{document}